\documentclass[reprint,aps,prd,groupedaddress,nofootinbib,superscriptaddress,eqsecnum,amsmath,amssymb]{revtex4}
\usepackage{amsfonts}
\usepackage{graphicx}
\usepackage{bm}
\usepackage{float}
\usepackage{url}
\usepackage[colorlinks=true,
    linkcolor=blue,
    citecolor=blue,
    urlcolor=blue]{hyperref}
\usepackage[figurename=Fig.]{caption}
    
\usepackage{lineno}
\usepackage{enumitem}

\begin{document}

\title{Revisiting Buchdahl transformations: New static and rotating black holes in vacuum,  double copy, and hairy extensions}

\author{Jos\'e Barrientos}
\email{jbarrientos@academicos.uta.cl}
\affiliation{Sede Esmeralda, Universidad de Tarapac\'a, Avenida Luis Emilio Recabarren 2477, Iquique, Chile}
\affiliation{Institute of Mathematics of the Czech Academy of Sciences, \v{Z}itn\'a 25, 115 67 Praha 1, Czech Republic}

\author{Adolfo Cisterna}
\email{adolfo.cisterna@mff.cuni.cz}
\affiliation{Sede Esmeralda, Universidad de Tarapac\'a, Avenida Luis Emilio Recabarren 2477, Iquique, Chile}
\affiliation{Institute of Theoretical Physics, Faculty of Mathematics and Physics,
Charles University, V Hole{\v s}ovi{\v c}k{\' a}ch 2, 180 00 Praha 8, Czech Republic}

\author{Mokhtar Hassaine}
\email{hassaine@inst-mat.utalca.cl}
\affiliation{Instituto de Matem\'{a}ticas,
Universidad de Talca, Casilla 747, Talca, Chile}

\author{Julio Oliva}
\email{juoliva@udec.cl}
\affiliation{Departamento de F\'{\i}sica, Universidad de Concepci\'{o}n, Casilla, 160-C,
Concepci\'{o}n, Chile}

\begin{abstract}
This paper investigates Buchdahl transformations within the framework of Einstein and Einstein-Scalar theories. Specifically, we establish that the recently proposed Schwarzschild-Levi-Civita spacetime can be obtained by means of a Buchdahl transformation of the Schwarschild metric along the spacelike Killing vector. The study extends Buchdahl's original theorem by combining it with the Kerr-Schild representation. In doing so, we construct new vacuum-rotating black holes in higher dimensions which can be viewed as the Levi-Civita extensions of the Myers-Perry geometries. Furthermore, it demonstrates that the double copy scheme within these new generated geometries still holds, providing an example of an algebraically general double copy framework. 
In the context of the Einstein-Scalar system, the paper extends the corresponding Buchdahl theorem to scenarios where a static vacuum seed configuration, transformed with respect to a spacelike Killing vector, generates a hairy black hole spacetime. We analyze the geometrical features of these spacetimes and investigate how a change of frame, via conformal transformations, leads to a new family of black hole spacetimes within the Einstein-Conformal-Scalar system.
\end{abstract}

\maketitle
\tableofcontents
\section{Introduction}

Einstein field equations represent a set of second-order, nonlinear coupled partial differential equations, of a mixed type since the momentum and Hamiltonian constraints are of elliptic type, while the evolution of the initial data is given by a hyperbolic system. Due to this inherent complexity, direct brute force integration of these equations proves challenging, with successful integration typically limited to cases exhibiting a high degree of symmetry. Consequently, the pursuit of exact solutions necessitates the utilization of specialized techniques, with Lie point symmetries serving as a prominent approach \cite{Stephani:2003tm}. These symmetries refer to a set of local transformations that consistently map each solution of the system to another solution within the same system. Hence, possessing a solution to the field equations and identifying a Lie point symmetry within the framework enables the discovery of additional solutions to the same equations. Although this seems to be a sort of recipe, there is a highly nontrivial effort behind the finding of a new Lie point symmetry, as they are usually hidden in the tensor language. 
Moreover, recognizing and accurately assessing the applicability of a particular Lie point symmetry is of paramount importance, as it may occur that the entirety of the symmetry's potential may go unexplored due to a lack of physical intuition.
In the realm of the electro-vacuum, the pioneering contributions of Ernst \cite{Ernst:1967wx,Ernst:1967by}, Geroch \cite{Geroch:1970nt}, and Kinnersley \cite{Kinn1}, building upon earlier works by Ehlers \cite{Ehlers:1957zz,Ehlers:1959aug}, and Harrison \cite{harrison1968new}, have played a pivotal role in advancing our understanding of exact solutions to the Einstein-Maxwell equations. Particularly significant is the Ernst scheme, which formulates the Einstein-Maxwell field equations in a space characterized by two complex potentials, one gravitational and the other electromagnetic, that enormously facilitates the detection of certain Lie point symmetries otherwise hidden in the Einstein-Maxwell theory, the so-called Ehlers and Harrison transformations \cite{Ehlers:1957zz,Ehlers:1959aug,harrison1968new}.

In this study, our objective is to reexamine and extend the discussion on arguably two of the earliest Lie point symmetries pertinent to the Einstein field equations. Specifically, we focus on the transformations introduced by Buchdahl \cite{Buchdahl:1956zz,Buchdahl:1959nk} and Janis-Robinson-Winicour (JRW) \cite{Janis:1969ivo}, which represent straightforward symmetries within the contexts of the Einstein vacuum, the Einstein-Scalar, and the Einstein-Scalar-Maxwell theories. These transformations facilitate the generation of new solutions from existing static vacuum solutions of the Einstein equations and have been very useful in the identification of novel spacetimes.
Buchdahl transformations are categorized into two types: first \cite{Buchdahl:1956zz} and second kind \cite{Buchdahl:1959nk}.\footnote{The categorization we have introduced here is new and serves the purpose of distinguishing between different types of transformations.} The former type converts any static vacuum solutions of the Einstein equations into a new ``reciprocal" static vacuum solution, while the latter type transforms any static vacuum solution into a static solution augmented with a nontrivial scalar field, specifically, a massless scalar field relevant to the Einstein-Scalar system. On the other hand, the JRW transformation \cite{Janis:1969ivo} generates an electrically charged solution from a known Buchdahl scalar solution, maintaining the scalar profile from the uncharged seed.
A Buchdahl transformation of the first kind may alter Minkowski spacetime, yielding a reciprocal spacetime that may not correspond to Minkowski spacetime in disguise. However, when applied to the Schwarzschild spacetime, it reproduces the Schwarzschild geometry in a different gauge, a result consistent with Birkhoff's theorem. In fact, Buchdahl transformations of the first kind (as well as Buchdahl transformations of the second kind), as typically employed, do not alter the spherical symmetry of the reciprocal solution, and in consequence, the latter cannot depart from the Schwarzschild spacetime.
Conversely, Buchdahl transformations of the second kind \cite{Buchdahl:1959nk} are recognized for their inability to introduce a scalar field onto Minkowski spacetime, but they do so onto the Schwarzschild geometry, producing the well-known Fisher-Janis-Newman-Winicour (FJNW) \cite{Fisher:1948yn,Janis:1968zz} spacetime. This spacetime, a solution of Einstein gravity enhanced with a massless minimally coupled scalar field, becomes electrically charged in the context of Maxwell theory through JRW transformations \cite{Janis:1969ivo}. While the scalar field profile remains unaltered from the uncharged seed, the electric configuration adopts a Coulombian nature.

In its original formulation \cite{Buchdahl:1956zz}, Buchdahl theorem (of the first kind) can be presented as follows: Consider any $d-$dimensional vacuum solution of the Einstein field equations 
\begin{equation}
    ds_0^2=g_{\mu\nu}dx^\mu dx^\nu=g_{aa}(x^k)(dx^a)^2+g_{ij}(x^k)dx^idx^j,
    \label{vacuum}
\end{equation}
which is said to be static with respect to a coordinate ``$a$", namely, that it satisfies $g_{ai}=0=\partial_a g_{\mu\nu}=0$. Here, $a$ represents a cyclic coordinate, and $i,j$ run from $1$ to $d-1$.\footnote{Note that the index $a$ is fixed, no summation is assumed even when repeated.}
It is proven that, knowing \eqref{vacuum},
a new static and vacuum solution ($d\geq4$) is automatically given by 
\begin{equation}\label{Buchdahl1}
    ds^2=(g_{aa})^{-1}(dx^a)^2+(g_{aa})^{\frac{2}{d-3}}g_{ij}dx^idx^j,
\end{equation}
which is said to be reciprocal to the original seed. In addition, it is proven \cite{Buchdahl:1959nk} (Buchdahl transformation of the second kind) that given a static vacuum solution of the form \eqref{vacuum}, the spacetime configuration ($d\geq4$)
\begin{equation}
\begin{aligned}
    ds^2_{ES}&=(g_{aa})^\beta(dx^a)^2+(g_{aa})^{\frac{1-\beta}{d-3}}g_{ij}dx^idx^j,\\
    \Phi&=\sqrt{\frac{(d-2)(1-\beta^2)}{4(d-3)\kappa}}\ln{(g_{aa})}, \label{Buchdahl2}
\end{aligned}
\end{equation}
will correspond to a solution of the Einstein-Scalar field equations 
\begin{equation}    R_{\mu\nu}=\kappa\partial_\mu\Phi\partial_\nu\Phi,\qquad \Box\Phi=0, \label{ESeq}
\end{equation}
where $\kappa=8\pi G$ represents Newton's constant, and $|\beta|<1$ a scalar hair parameter.

The conventional understanding of staticity is tied to the metric's time independence and the absence of stationary terms in the metric tensor. Consequently, investigations into the Buchdahl theorem have primarily occurred within the framework where the cyclic coordinate $a$ represents the time coordinate. In this context, this is what we previously called ``the typical" manner of applying the Buchdahl transformations \eqref{Buchdahl1} and \eqref{Buchdahl2}.
However, this limitation is not inherent in the theorem itself; rather, $a$ could denote any cyclic coordinate associated with any other Killing vector of the geometry, not necessarily the timelike Killing vector $\partial_t$. An intriguing avenue of inquiry involves employing Buchdahl transformations concerning spacelike Killing vectors. Therefore, a logical progression of exploration involves examining the effects of a Buchdahl transformation when applied along the Killing vector $\partial_\phi$. 

In a recent publication \cite{Mazharimousavi:2024hrg}, a vacuum solution to the Einstein field equations has been introduced, termed the Schwarzschild-Levi-Civita black hole.
This geometry depicts a Schwarzschild black hole situated within a Levi-Civita spacetime, as its geometric analysis can testify. The geometry corresponds to the Levi-Civita spacetime in the massless limit, while it asymptotically approaches a Levi-Civita spacetime at radial infinity, or in terms of cylindrical coordinates, asymptotically far from the symmetry axis. Although not advertised in \cite{Mazharimousavi:2024hrg}, further analysis reveals that this solution is algebraically general, namely, of Petrov type I, indicating that its classification departs from the general expanding Pleban\'ski-Demia\'nski class \cite{Plebanski:1976gy}, where most known black holes belong to.
Furthermore, particularly noteworthy reveals to be the fact that this solution corresponds to a Buchdahl-transformed version of the Schwarzschild black hole. However, unlike previous transformations, which were typically applied with respect to the timelike Killing vector $\partial_t$,
this transformation occurs concerning the azimuthal spacelike Killing vector $\partial_\phi$. 

The Schwarzschild-Levi-Civita spacetime prompts a reevaluation of Buchdahl's transformations. Specifically, we aim to extend the Buchdahl theorem to encompass geometries featuring two Killing vectors in dimension four. However, an extension with 
an arbitrary number of commuting Killing vectors and in an arbitrary number of dimensions is also feasible. We will examine the new geometries achievable via this generalized Buchdahl theorem when applied to spacelike Killing vectors. Furthermore, it will be demonstrated that the solution introduced in \cite{Mazharimousavi:2024hrg} admits a Kerr-Schild representation, with the Kerr-Schild vector being null and geodesic but not shear-free. This observation aligns with the Goldberg-Sachs theorem \cite{GS}, which stipulates that the geometry must belong to algebraically special spacetimes for a null and shear-free geodesic congruence.
This naturally leads to the question of how 
Buchdahl transformations can be applied to rotating vacuum spacetimes. Explicit Buchdahl versions of the Myers-Perry black hole \cite{Myers:1986un} and their Kerr-Schild form will be provided. In addition, it will be clearly seen why four-dimensional rotating black holes cannot be subjected to Buchdahl transformations.  
Additionally, we will elucidate how the double copy scheme \cite{Bern:2008qj,Bern:2010ue,Monteiro:2014cda}, which establishes connections between solutions of Einstein-Maxwell theory and Maxwell theory in flat spacetime, operates in the presence of Buchdahl-transformed geometries.
Lastly, at a static level, we will revisit the standard Buchdahl theorem within Einstein-Scalar theory \cite{Buchdahl:1959nk}, in order to construct new hairy extensions of the Schwarzschild-Levi-Civita black hole. In addition, we will 
introduce a framework through which an external magnetic field can influence these solutions, providing a magnetic generalization of the JRW transformations \cite{Janis:1969ivo}.
We will analyze the main geometric properties of these new hairy black hole solutions and provide several avenues to be explored in the context of black holes in scalar-tensor theories. 

This work is structured as follows: In Section \ref{sec2}, we delve into the main properties of the Levi-Civita black hole \cite{Mazharimousavi:2024hrg}, elucidating its evident connection with the Buchdahl theorem \eqref{Buchdahl1} and providing a magnetized version that regularizes the curvature singularity it presents along the symmetry axis.
Section \ref{sec3} extends the Buchdahl theorem in vacuum in several directions. First, 
to scenarios where two commuting Killing vectors of a four-dimensional geometry are considered. Next, Buchdahl and Kerr-Schild transformations are composed, resulting in the Kerr-Schild form of the Schwarzschild-Levi-Civita black hole and facilitating the construction of new higher-dimensional rotating black holes. In particular, Buchdahl-transformed extensions of the Myers-Perry spacetime. In addition, leveraging the Kerr-Schild form of our solutions, we establish a novel double copy scheme, notable for its algebraically general nature. Section \ref{sec4}  presents a generalization of the original Buchdahl theorem within the Einstein-Scalar system, facilitating the construction of a Levi-Civita version of the Fisher-Janis-Newman-Winicour black hole \cite{Fisher:1948yn,Janis:1968zz,Janis:1969ivo}. We analyze its key geometric features and introduce its magnetically charged extension, in an attempt to generalize the standard JRW electric transformation \cite{Janis:1969ivo}.
Additionally, we perform the well-known Bekenstein change of frame \cite{Bekenstein:1974sf}, obtaining the Levi-Civita extension of the celebrated BBMB black hole \cite{Bocharova:1970skc,Bekenstein:1974sf}.
Finally, in Section \ref{sec5}, we conclude by outlining further avenues for exploration within the realm of these novel geometries.

\section{Revisiting the Schwarzschild-Levi-Civita Black Hole}\label{sec2}

As we have already pointed out the Schwarzschild-Levi-Civita black hole \cite{Mazharimousavi:2024hrg}, although it can be easily integrated from the field equations, corresponds with the Buchdahl-transformed version of the Schwarzschild geometry, however, not with respect to the standard timelike Killing vector but with respect to the spacelike Killing vector $\partial_\varphi$. As presented in \cite{Mazharimousavi:2024hrg} the geometry reads 
\begin{equation}
    ds^2=\frac{d\varphi^2}{\lambda_0^2r^2\sin^2\theta}+\lambda_0^2r^4\sin^4\theta\left[-\left(1-\frac{2M}{r}\right)dt^2+\frac{dr^2}{\left(1-\frac{2M}{r}\right)}+r^2 d\theta^2\right], \label{LVBH}
\end{equation}
being $\lambda_0$ an integration constant and $M$ the mass parameter. 
In fact, considering the coordinate $a$ to be the azimuthal angle $\varphi$ and remaining in four dimensions, the Buchdahl-transformed metric \eqref{Buchdahl1} becomes 
\begin{equation}
ds^2=\frac{d\varphi^2}{r^2\sin^2\theta}+r^4\sin^4\theta g_{ij}dx^idx^j.
\end{equation}
As a matter of fact, if the seed spacetime is chosen to be the Schwarzschild geometry, this precisely reproduces the Schwarzschild-Levi-Civita spacetime \eqref{LVBH} for $\lambda_0=1$. Contrary to the case in which the timelike Killing vector has been considered in the transformation, the resulting metric cannot be turned back to the seed by a large diffeomorphism as the initial spherical symmetry of the seed has been lost. 
Direct inspection reveals that the effect of the aforementioned Buchdahl transformation consists of embedding the given seed on a Levi-Civita spacetime \cite{LC}, {namely, a static, cylindrically symmetric vacuum of Einstein equations}. Actually, considering the Minkowski spacetime in cylindrical coordinates $(t,\rho,z,\varphi)$
\begin{equation}
    ds_0^2=-dt^2+d\rho^2+dz^2+\rho^2d\varphi^2,
\end{equation}
making use of \eqref{Buchdahl1} with respect to the azimuthal coordinate provides us with 
\begin{equation}
    ds^2=\rho^4(-dt^2+d\rho^2+dz^2)+\frac{d\varphi^2}{\rho^2}.
\end{equation}
This spacetime is nothing else than the Levi-Civita spacetime
\begin{equation}
 ds_{LC}^2=-\rho^{4\sigma}dt^2+k^2\rho^{4\sigma(2\sigma-1)}(d\rho^2+dz^2)+\rho^{2(1-2\sigma)}d\varphi^2,
\end{equation}
with mass per unit length $\sigma=1$ and $k=1$. Notice that a rescaling of the noncompact coordinates is necessary. This represents a particular case of vacuum Weyl spacetimes in which the geometry features an additional spacelike Killing vector, $\partial_z$. Now, the Schwarzschild black hole embedded in the Levi-Civita spacetime certainly behaves asymptotically as the Levi-Civita spacetime and it acquires most of its features, besides losing the evident symmetry with respect to the $z$-coordinate which is lost due to the presence of the mass term $M$. 
{Therefore, it is of a general algebraic nature, namely, of Petrov type I. Its background, the Levi-Civita spacetime, is algebraically special for $\sigma=(-1/2,1/4,1)$ and conformally flat for $\sigma=(0,1/2,\infty)$. 
Moreover, since it belongs to a vacuum Weyl solution, its interpretation in terms of a Newtonian potential is direct to obtain, representing the gravitational potential of an infinite uniform line source with a mass per unit length expressed by $\sigma$. Consequently, the Schwarzschild-Levi-Civita black hole will always feature a curvature singularity all over the symmetry axis, which will be nothing but the relativistic analog of the Newtonian line source sourcing the Levi-Civita spacetime.}   

The curvature singularity located at the axis of symmetry of the Schwarzschild-Levi-Civita black hole can be cured by embedding the geometry on an external magnetic field. In direct analogy with the Schwarzschild-Melvin spacetime \cite{Bonnor_1954,Melvin1966}, which is obtained via a magnetic Harrison transformation \cite{harrison1968new}, we construct the Melvin extension of the Schwarzschild-Levi-Civita black hole 
\begin{equation}\label{magneticbuchdahl}
\begin{aligned}
ds^2&=\Lambda^2\left(r^4\sin^4\theta\right)\left[-\left(1-\frac{2M}{r}\right)dt^2+\frac{dr^2}{1-\frac{2M}{r}}+r^2d\theta^2\right]+\frac{1}{(r^2\sin^2\theta)\Lambda^2} d\varphi^2,\\
A&=\frac{B}{\Lambda}\left(\frac{1}{r^2\sin^2\theta}\right) d\varphi,\\
\end{aligned}
\end{equation}
where $\Lambda=1+\frac{B^2}{r^2\sin^2\theta}$. The metric exhibits the standard curvature singularity at $r=0$ only. As expected, asymptotically far from the axis of symmetry it behaves as a Levi-Civita spacetime. 
 {Interestingly enough, close to the axis of symmetry, it is possible to show that 
\begin{equation}
\begin{aligned}
ds^2&\underset{r\to 0}{\sim}B^4\left[\frac{2M}{r}dt^2-\frac{r}{2M}dr^2+r^2(d\theta^2+\sin^2\theta d\varphi^2)\right],\\
A&\underset{r\to 0}{\sim}B^3d\varphi.
\end{aligned}
\end{equation}
Redefining the azimuthal coordinate as  $\varphi\rightarrow B^4\varphi$, and rescaling the line element by a constant conformal factor $B^4$, the spacetime is given by nothing else than the Schwarzschild metric, with a pure gauge Maxwell field.}

\section{Extending Buchdahl theorem in vacuum} \label{sec3}
Thus far, our investigation has yielded a preliminary observation: employing the Buchdahl transformations relative to the spacelike Killing vector $\partial_\varphi$ presents an intriguing approach for generating novel spacetimes within the framework of Einstein's theory. As shown below in  Sec. \ref{sec4}, this method also extends to the construction of distinctive spacetimes within Einstein-Scalar and Einstein-Scalar-Maxwell theories. The primary consequence appears to be the incorporation of the original seed onto Levi-Civita spacetime, at least when employing one single transformation. This significantly modifies the geometric characteristics of these newly derived solutions compared to the familiar solutions obtained through the application of Buchdahl transformations according to the standard lore.

\subsection{Buchdahl theorem and its extension in \texorpdfstring{$d=4$}{Lg} with two Killing vectors}

A generalization of the Buchdahl transformation can be achieved, when the number of Killing vectors is greater or equal than 2, through compositions between two different transformations associated with two different Killing vectors. We will initially limit ourselves to the case in four dimensions with exactly 2 Killing vectors, however, this construction can be extended in arbitrary dimensions and with any number of Killing vectors.

Hence, let $g_{\mu\nu}$ be the metric tensor in four dimensions with coordinates defined by $x^{\mu}=(t, x^i,\varphi)$ and let $\xi^{\mu}\partial_{\mu}=\partial_{t}$ and $\chi^{\mu}\partial_{\mu}=\partial_{\varphi}$ be two  Killing vector fields, timelike and spacelike, respectively. In addition, let us assume that $g_{it}=g_{i\varphi}=g_{t\varphi}=0$, that is, the metric tensor acquires the form  $g=\left(-g_{tt},\,g_{ij}, \,g_{\varphi\varphi}\right)$ with all the metric components being independent of $t$ and $\varphi$.\\
In order to generalize the original Buchdahl theorem (in four dimensions) to the case in which the transformation involves the  two Killing vectors, we start by defining the family of applications $T_{(n,m)}$, with $n, m\in \mathbb{Z}$, that acts on the metric components as follows 
\begin{equation}
T_{(n,m)}(g)=\left[-\left(g_{tt}\right)^{a_{n,m}}\left(g_{\varphi\varphi}\right)^{a_{m,n}}g_{tt},\,
\left(g_{tt}\right)^{b_{n,m}}\left(g_{\varphi\varphi}\right)^{b_{-n,-m}}g_{ij},\,
\left(g_{tt}\right)^{-a_{n,m}}\left(g_{\varphi\varphi}\right)^{-a_{m,n}}g_{\varphi\varphi}\right],
\label{Buchdahlgener}
\end{equation}
where we have defined
\begin{eqnarray}
\label{defab}
a_{n,m}=2(-1)^{n+m}n,\qquad b_{n,m}=(n+m)(n+m+1).
\end{eqnarray}
It can be proven that if $g_{\mu\nu}=\left(-g_{tt},\,g_{ij}, \,g_{\varphi\varphi}\right)$ is a vacuum metric, the reciprocal spacetime metrics defined by
\begin{subequations}
\label{bg}
\begin{eqnarray}
\label{bg1}
&&g^{\prime}=T_{(n,n)}(g):=T^{(1)}_n(g),\\
\label{bg2}
&&g^{\prime\prime}=T_{(n,n-1)}(g):=T^{(2)}_n(g),
\end{eqnarray}
\end{subequations}
will also satisfy $R_{\mu\nu}=0$, and this for any integer $n\in \mathbb{Z}$. 

Many comments can be made concerning this result. First of all, note that since $a_{n,m}$ and $b_{n,m}$ (\ref{defab}) are always even, the signature of the transformed metric  $T_{(n,m)}(g)$ is the same as the original one, $g$. One can also notice that for the Killing vector field $\xi$ (resp. $\chi$), the standard Buchdahl transformation as defined in \cite{Buchdahl:1956zz} will correspond to (\ref{bg2}) with $n=1$ (resp. $n=0$). For example, for a starting metric given by the 
Schwarzschild solution with coordinates $(t,r,\theta,\varphi)$ and metric components $g_{tt}=g_{rr}^{-1}=1-\frac{2M}{r}$, $g_{\theta\theta}=r^2$ and $g_{\varphi\varphi}=r^2\sin^2\theta$, the transformation  (\ref{bg2}) with $n=0$ yields the Schwarzschild-Levi-Civita black hole (\ref{LVBH}), recently presented in \cite{Mazharimousavi:2024hrg}. Note that for the transformed Schwarzschild solution with $n=1$, the resulting metric is nothing but 
the Schwarzschild spacetime itself written in a different set of coordinates, see below (\ref{pv223}). It is also interesting to see that the set of all the transformations (\ref{bg}) forms a group under composition 
\begin{eqnarray*}
&&T^{(1)}_0=\mbox{id},\qquad (T^{(1)}_n)^{-1}=T^{(1)}_{-n},\qquad (T^{(2)}_n)^{-1}=T^{(2)}_{n}\\
&&T^{(1)}_m\circ T^{(1)}_n=T^{(1)}_{n+m},\quad T^{(2)}_m\circ T^{(2)}_n=T^{(1)}_{n-m},\quad T^{(2)}_m\circ T^{(1)}_n=T^{(2)}_{n+m},\qquad T^{(1)}_m\circ T^{(2)}_n=T^{(2)}_{n-m}.
\end{eqnarray*}
Moreover, as previously stated, the transformations (\ref{bg}) are nothing but the compositions of the standard Buchdahl transformations associated with each of the two Killing vectors. In other words, the group of generalized Buchdahl transformations are generated by $T^{(2)}_{1}$ and $T^{(2)}_{0}$ only, which are exactly the standard Buchdahl transformations associated with $\xi$ and $\chi$. That is, any transformation (\ref{bg}), can be expressed as an alternated composition of the transformations $T^{(2)}_{1}$ and $T^{(2)}_{0}$ as
\begin{equation}
{\cal T}_1:=T^{(2)}_{1}\circ T^{(2)}_{0}\circ T^{(2)}_{1}\circ T^{(2)}_{0}\cdots \qquad \mbox{or}\qquad {\cal T}_0:=T^{(2)}_{0}\circ T^{(2)}_{1}\circ T^{(2)}_{0}\circ T^{(2)}_{1}\cdots.
\end{equation}
More precisely, ${\cal T}_1$ will generate all the transformations $T^{(1)}_{-n}$ and $T^{(2)}_{n}$ for $n\in \mathbb{N}$ while ${\cal T}_0$ will induce the remaining transformations, namely $T^{(1)}_{n}$ and $T^{(2)}_{-n}$. For example, one can represent the transformation $T^{(1)}_3$ as 
\begin{equation}
T^{(1)}_3=T^{(2)}_{0}\circ T^{(2)}_{1}\circ T^{(2)}_{0}\circ T^{(2)}_{1}\circ T^{(2)}_{0}\circ T^{(2)}_{1}.
\end{equation}

In order to be concrete, once again we exemplify with the Schwarzschild-Levi-Civita black hole of \cite{Mazharimousavi:2024hrg}
\begin{eqnarray}
ds_{SLC}^2=r^4\sin^4\theta\left[-\left(1-\frac{2M}{r}\right)dt^2+\frac{dr^2}{1-\frac{2M}{r}}+r^2d\theta^2\right]+\frac{d\varphi^2}{r^2\sin^2\theta}.
\label{lv2}
\end{eqnarray}
As already mentioned, this spacetime can be obtained from the Schwarzschild metric
\begin{eqnarray}
d{s}_{Schw}^2=-\left(1-\frac{2M}{r}\right)dt^2+\frac{dr^2}{1-\frac{2M}{r}}+r^2d\theta^2+r^2\sin^2\theta d\varphi^2,
\label{Schwa}
\end{eqnarray}
by means of the transformation (\ref{bg2}) with $n=0$, that is  $g_{SLC}=T^{(2)}_{(0,-1)}g_{Schw}$. Let us now go beyond a single transformation, and from the vacuum solution (\ref{lv2}), operate with a generalized transformation with the timelike vector field $\xi^{\mu}\partial_{\mu}=\partial_{t}$. The resulting new vacuum spacetime, dubbed extended Schwarzschild-Levi-Civita, reads  
\begin{align}
ds_{ESLC}^2=\frac{-dt^2}{r^4\sin^4\theta\left(1-\frac{2M}{r}\right)}+\left[r^4\sin^4\theta \left(1-\frac{2M}{r}\right)\right]^2\left[r^4 \sin^4\theta\left(\frac{dr^2}{1-\frac{2M}{r}}+r^2d\theta^2\right)+\frac{d\varphi^2}{r^2\sin^2\theta}\right],
\label{lv22}
\end{align}
and is simply given via the transformations 
and  $g_{ESLC}=T^{(2)}_{(1,0)}g_{SLC}=T^{(2)}_{(1,0)}\circ T^{(2)}_{(0,-1)}g_{Schw}=T^{(1)}_{(-1,-1)}g_{Schw}$. {This spacetime has an asymptotic region as $r\rightarrow\infty$ which is locally flat, namely $R^{\mu\nu}_{\ \ \alpha\beta}
$ approaches zero there, and even more the Killing vector $\partial_t$ becomes null, for arbitrary large $r$. The interior of the spacetime ends at the singularity $r=2M$, which can be seen by computing the Kretchmann scalar, that diverges at the axes $\theta=0,\pi$, as well.} In a similar way, we can start by acting on the Schwarzschild metric with a standard Buchdahl transformation with respect to the timelike Killing vector $\partial_t$. This metric, which is nothing else than the Schwarzschild metric in a different set of coordinates, can be obtained via (\ref{bg2}) with $n=1$ as $\bar{g}_{Schw}=T^{(2)}_{(1,0)}g_{Schw}$. It is given by 
\begin{equation}
d\bar{s}_{Schw}^2=\frac{-dt^2}{\left(1-\frac{2M}{r}\right)}+\left(1-\frac{2M}{r}\right)^2\Bigg[\frac{dr^2}{\left(1-\frac{2M}{r}\right)}+r^2d\theta^2+r^2\sin^2\theta d\varphi^2\Bigg].
\label{pv223}
\end{equation}
Acting on this metric with a Buchdahl transformation with respect to the spacelike Killing vector $\partial_\varphi$ provides the spacetime 
$\bar{g}_{ESLC}=T^{(2)}_{(0,-1)}\bar{g}_{Schw}=T^{(1)}_{(1,1)}g_{Schw}$ of which the line element reads 
\begin{align}
d\bar{s}_{ESLC}^2&=\Bigg[r^2\sin^2\theta\left(1-\frac{2M}{r}\right)^2\Bigg]^2\Bigg[-\frac{dt^2}{1-\frac{2M}{r}}+\left(1-\frac{2M}{r}\right)^2\left(\frac{dr^2}{(1-\frac{2M}{r})}+r^2d\theta^2\right)\Bigg]+\frac{d\varphi^2}{r^2\sin^2\theta \left(1-\frac{2M}{r}\right)^2}.
\label{lv223}
\end{align}
Note that this metric is not diffeomorphic to (\ref{lv22}). Therefore, the different vacuum metrics generated by means of these transformations (\ref{bg}) are not diffeomorphic in general, and this is consistent with the non-Abelian nature of the transformations. {The latter geometry \eqref{lv223}, also has an asymptotically locally flat region as $r\rightarrow\infty$, and the spacetime interior extends up to $r=2M$, where the spacetime is singular. In this case the Killing field $\partial_t$ remains timelike for $2M<r<\infty$.}

Finally, we propose to re-derive the previous results with a static and spherical seed metric using the mini-superspace formalism. This approach also presents the advantage of specifying the thermodynamic properties of the solutions obtained by means of the generating technique method. As usual, we work in the Euclidean formalism with the Euclidean time $\tau$ related to the Lorentzian time $t$ by $t=i\tau$. In order to fit with the prescription of (\ref{Buchdahlgener}), we shall consider an Euclidean ansatz of the form
\begin{equation}
\begin{aligned}
ds_E^2=\left(N^2 f\right)^{a_{n,n-1}}\Lambda^{a_{n-1,n}}\, d\tau^2+\left(N^2 f\right)^{b_{n,n-1}}\Lambda^{b{-n,1-n}}\left[\frac{dr^2}{f}+r^2d\theta^2\right]+\left(N^2 f\right)^{a_{-n,1-n}}\Lambda^{-b_{-n,1-n}}\, r^2\sin^2\theta d\varphi^2,
\end{aligned}
\end{equation}
where $f=f(r)$, $N=N(r)$ and $\Lambda=\Lambda(r,\theta)$, and where the factors $a$ and $b$ are defined in (\ref{defab}). The Euclidean Einstein-Hilbert action evaluated on this ansatz (after some integration by parts) and with a periodic time $\tau$ of period $\beta=\frac{1}{T}$, with $T$ being the temperature,  reads 
\begin{equation}
\begin{aligned}
\label{reduaction}
I_E&=\frac{1}{2\kappa}\int \sqrt{g_E}\,d^4x\,R,\\
&=\frac{\beta}{\kappa}\int_{\theta} \int_{r} N\Bigg[\sin\theta\left(rf^{\prime}+f-1+\left(\frac{\partial_{\theta}\Lambda}{\Lambda}\right)^2+r^2 f\left(\frac{\Lambda'}{\Lambda}\right)^2-2rf\frac{\Lambda'}{\Lambda}\right)-2\cos\theta\frac{\partial_{\theta}\Lambda}{\Lambda}\Bigg].\nonumber
\end{aligned}
\end{equation}
It is interesting to note that the reduced action (\ref{reduaction}) is generic and does not depend on the coefficients $a$ and $b$, and its field equations obtained by varying with respect to $f$, $N$, and $\Lambda$ are easily integrated. Indeed, one can show that the equation we obtain via variation with respect to $f$ can be re-written as 
$$
\frac{N'}{N}=r\left(\frac{\Lambda'}{\Lambda}\right)^2-\frac{2\Lambda'}{\Lambda},
$$
which automatically implies that $\Lambda$ is separable in product as $\Lambda=r^2 H(\theta)$, and consequently $N=\mbox{cte}$. Finally, the integration of the remaining equations yields
\begin{eqnarray}
\label{solE}
f(r)=1-\frac{2M}{r},\qquad\qquad N=\mbox{cte},\qquad\qquad \Lambda(r,\theta)=r^2\sin^2\theta.
\end{eqnarray}
This result is in complete accordance with the extended Buchdahl transformations (\ref{Buchdahlgener}) together with the Birkhoff theorem. The thermodynamics can be carried out by adding a boundary term $B_E$ to the Euclidean action such that the full variation vanishes $\delta(I_E+B_E)=0$. In our case, the variation of the appropriate boundary term is given by
$$
\delta B_E=-\frac{\beta}{\kappa}\int_{\theta=0}^{\pi}\int_{r=r_h}^{\infty} N\Big[\sin\theta\Big(r\delta f+2\delta\ln\Lambda(1+r^2f-2rf)\Big)-2\cos\theta\delta\ln\Lambda\Big],
$$
where $r_h=2M$ is the location of the horizon. For the class of solutions (\ref{solE}), since $\delta\ln\Lambda=0$, we end up with the standard Schwarzschild boundary term
$$
B_E=\beta\left(\frac{4M}{\kappa}\right)-\frac{4\pi r_h^2}{\kappa},
$$
from which one can infer the mass ${\cal M}=\frac{4M}{\kappa}$, and that the entropy satisfies the one-quarter area law ${\cal S}=\frac{4\pi r_h^2}{\kappa}$.

As a last comment before closing this section, we recall that the Buchdahl transformation of the Schwarzschild solution along the timelike Killing vector yields, after an appropriate re-definition of the radial coordinate, the Schwarzschild spacetime. There is nevertheless an example where this transformation on a vacuum black hole solution generates a new solution which is not diffeomorphic to the starting seed. Indeed, in \cite{Hassaine:2015ifa}, a five-dimensional  vacuum solution with a three-dimensional Nil geometry horizon was constructed, of which the line element is given by
\begin{equation}
ds^2=-\left(1-\frac{M}{r}\right)dt^2+\frac{dr^2}{r^5\left(1-\frac{M}{r}\right)}+\frac{1}{r}(dx_1^2+dx_2^2)+r(dx_3-x_1 dx_2)^2.
\end{equation}
It is clear that the Buchdahl transformation can only be done along the timelike Killing vector field yielding to a new vacuum solution with a Nil geometry horizon
\begin{equation}
ds^2=-\frac{dt^2}{\left(1-\frac{M}{r}\right)}+\left(1-\frac{M}{r}\right)\Bigg[\frac{dr^2}{r^5\left(1-\frac{M}{r}\right)}+\frac{1}{r}(dx_1^2+dx_2^2)+r(dx_3-x_1 dx_2)^2\Bigg],
\end{equation}
which after the change {$r=-\rho+M$, and then performing the redefinition $\rho=-r$ and $M\rightarrow-M$ leads to}
\begin{equation}
ds^2=-\left(1-\frac{M}{r}\right)dt^2+\frac{dr^2}{r^5\left(1-\frac{M}{r}\right)^5}+\frac{r}{(r-M)^2}(dx_1^2+dx_2^2)+r(dx_3-x_1 dx_2)^2.
\end{equation}
{In this case the range of the radial coordinate can be extended from $0<r<\infty$. The spacetime is singular at both ends, nevertheless, the null surface $r=M$ is a Killing horizon for the Killing field $\partial_t$, which is timelike for $r>M$ and spacelike when $r<M$. The surfaces at constant $t$ and constant $r$, are Nil geometries, leading to the following Killing vectors for the full spacetime $\left(\mathbf{e_1}=\partial_t,\mathbf{e_2}=2\left(x_1\partial_{x_2}-x_2\partial_{x_1}\right)+\left(x_1^2-x_2^2\right)\partial_z,\mathbf{e_3}=\partial_{x_1}+x_2\partial_{x_3},\mathbf{e_4}=\partial_{x_2},\mathbf{e_5}=\partial_{x_3}\right)$, which span the following solvable algebra
\begin{equation}
[\mathbf{e_2},\mathbf{e_3}]=-2\mathbf{e_4}\ , \ [\mathbf{e_2},\mathbf{e_4}]=2\mathbf{e_3} \ , \ [\mathbf{e_3},\mathbf{e_4}]=-\mathbf{e_5}\ .
\end{equation}
A further exploration of the geodesic motion on this spacetime, exploiting the isometry algebra, will be presented elsewhere.}

\subsection{Composition of Buchdahl and Kerr-Schild transformations}

In order to explore whether any of the spacetimes acquired through Buchdahl transformations can be expressed in a Kerr-Schild form, we delve into the composition of Buchdahl and Kerr-Schild transformations. Here, we will only use the Buchdahl transformation with respect to the spacelike 
Killing vector $\partial_\varphi$. This is due to the fact that the effect of the transformation with respect to the timelike Killing vector is pure gauge if the seed is either standard Minkowski or the Schwarzschild spacetime. Starting from this observation, we will then split the $d$-dimensional coordinates appropriately as $x^{\mu}=(x^i, \varphi)$, where $i=1,\ldots,d-1$. Our objective is to fix the conditions for which a metric written in the Kerr-Schild form maintains a Kerr-Schild decomposition under the action of a Buchdahl transformation with respect to the spacetime Killing vector $\partial_\varphi$. Let us assume that the starting vacuum metric admits a Kerr-Schild representation 
\begin{equation}
ds^2=ds_0^2+H(x^{\mu})l\otimes l,\label{ksm1}
\end{equation}
where the seed metric is $g_0$, and $l$ is a null and geodesic vector field for the seed and the full metrics
\begin{equation}
g^{{(0)}\mu\nu}l_{\mu}l_{\nu}=l^{\mu}(\nabla^{(0)}_{\mu}l_{\nu})=0=g^{\mu\nu}l_{\mu}l_{\nu}=l^{\mu}(\nabla_{\mu}l_{\nu})=0.
\end{equation}
Now, in order for the full metric to  satisfy the Buchdahl  conditions along the spacelike Killing vector $\partial_\varphi$, the seed metric $g_0$ must have the following decomposition
\begin{equation}
ds_0^2=g_{ij}^{(0)}dx^idx^j+g_{\varphi\varphi}^{(0)} d\varphi^2,
\end{equation}
while $H=H(x^i)$ and the component of the null $1-$form $l$ along the basis element $d\varphi$ is vanishing, i.e., $l=a_i(x^j) dx^i$. Under these conditions, it is clear that using the Buchdahl transformation, the metric defined by
\begin{equation}
d\tilde{s}^2= \left(g_{\varphi\varphi}^{(0)}\right)^{\frac{2}{d-3}} \left[g_{ij}^{(0)}dx^idx^j\right]+\frac{ d\varphi^2}{g_{\varphi\varphi}^{(0)}} + \left(g_{\varphi\varphi}^{(0)}\right)^{\frac{2}{d-3}} H(x^i)l\otimes l,
\label{ksm2}
\end{equation}
is also a vacuum metric. It remains to show that (\ref {ksm2}) has a Kerr-Schild representation, that is $l$ is null and geodesic with respect to seed and the full metric (\ref{ksm2}). In fact,  under our assumption, particularly that $l=a_i(x^j) dx^i$, the null character of $l$ is trivial while the geodesic equation simplifies to
\begin{equation}
l^{\mu}(\tilde{\nabla}_{\mu}l_{\nu})=l^i\left[\Gamma^{(0)j}_{ik}-\tilde{\Gamma}^{j}_{ik}\right]l_j=-\frac{1}{\Omega}\left[(\partial_i\Omega)\,\delta^j_k+(\partial_k\Omega)\, \delta^j_i-(\partial^j\Omega)\, g_{ik}\right]l^il_j,
\end{equation}
where $\Omega=\left(g_{\varphi\varphi}^{(0)}\right)^{\frac{1}{d-3}}$. By utilizing the null nature of the congruence $\tilde{g}^{\mu\nu}l_{\mu}l_{\nu}=l_i l^i=0$, it becomes evident that $l$ is geodesic as well.

Let us exemplify this outcome with two cases. We will first show that the Schwarzschild-Levi-Civita black hole can fit in the Kerr-Schild representation, and that this is related to the fact that the  Schwarzschild metric itself has a Kerr-Schild form respecting the previous conditions. Unfortunately, although Kerr's solution has a Kerr-Schild type representation, the latter is not compatible with the above-mentioned conditions unless the rotation parameter vanishes. However, in higher dimensions where there is more than one rotation parameter, one can show that the general Myers-Perry rotating black hole solutions with one vanishing rotation are in complete adequacy with the previous hypotheses and thus new spinning solutions can be generated.

\subsubsection{Kerr-Schild form of the Schwarzschild-Levi-Civita Black Hole}

Firstly, observe that the Schwarzschild-Levi-Civita black hole metric (\ref{lv2}) can be expressed through a Kerr-Schild transformation of the form
\begin{eqnarray}
ds_{SLC}^2=r^4\sin^4\theta\left[-dt^2+dr^2+r^2d\theta^2\right]+\frac{d\varphi^2}{r^2\sin^2\theta}+\left(2Mr^3\sin^4\theta\right)(dt+dr)^2,
\label{lv}
\end{eqnarray}
where $l=dt+dr$ defines a null and geodesic but not shearfree congruence, and that its standard form (\ref{lv2}) is recovered by implementing the following change of coordinates 
\begin{eqnarray}
dt\to dt+\frac{2M}{r-2M}dr.
\label{cge}
\end{eqnarray}
On the other hand, it is known that the Kerr-Schild representation of the Schwarzschild metric is given by
\begin{eqnarray}
ds^2_{Schw}=-dt^2+dr^2+r^2d\theta^2+r^2\sin^2\theta d\varphi^2+\left(\frac{2M}{r}\right)(dt+dr)^2,
\label{schw}
\end{eqnarray}
and that its standard form is recovered using the same change of variables (\ref{cge}). It is direct to observe that the Kerr-Schild representation of the Schwarzschild-Levi-Civita black hole (\ref{lv}) can be obtained from (\ref{schw}) using the transformation (\ref{ksm2}). Although this observation proceeds straightforwardly, it suggests an interesting fact. Indeed, it seems that the Schwarzschild-Levi-Civita black hole constitutes the first example of an algebraically general spacetime in vacuum admitting a Kerr-Schild representation. 
Note that in contrast to the Schwarzschild-Levi-Civita black hole (\ref{lv}), the metric (\ref{pv223}) cannot be obtained from a standard Kerr-Schild transformation but instead from a Kerr-Schild transformation in the $(t,r)-$sector and a conformal transformation in the orthogonal sector $(\theta,\varphi)$. Indeed, the metric defined by
\begin{equation}
d\bar{s}_{Schw}^2=-dt^2+dr^2-\left(\frac{2M}{r-2M}\right)\left(dt+dr\right)^2+\left(1-\frac{2M}{r}\right)^2 \left[r^2d\theta^2+r^2\sin^2\theta d\varphi^2\right],
\label{tgks}
\end{equation}
following the change of variable, $dt\to dt-\frac{2M}{r}dr$, yields (\ref{pv223}).

\subsubsection{A rotating example: Myers-Perry black holes \textit{\`a la} Buchdahl}

The next example is provided by the rotating Myers-Perry vacuum solutions \cite{Myers:1986un} in $d=5$, and later generalized in any higher dimension $d\geq 5$.  In a $d-$dimensional Myers-Perry solution, the rotation group is $\mbox{SO}(d-2)$ which has $[\frac{d-1}{2}]$ independent Casimir invariants corresponding to the rotations in the distinct planes (each of them associated to a different rotation parameter). One can show that for the Myers-Perry solutions the hypothesis allowing the Buchdahl transformation is not respected unless (and at least) one of the rotation parameters vanishes. We shall also take advantage from the fact that the rotating Myers-Perry solutions admit a Kerr-Schild representation to present the new generated spinning solutions in the Kerr-Schild form.

In five dimensions, the Myers-Perry solution with two rotations has the following Kerr-Schild representation 
\begin{equation}
ds^2_{MP}=-dt^2+\frac{r^2\rho^2}{(r^2+a_1^2)(r^2+a_2^2)}dr^2+\rho^2d\theta^2+(r^2+a_1^2)\sin^2\theta d\varphi_1^2+(r^2+a_2^2)\cos^2\theta d\varphi_2^2-\frac{M}{\rho^2}\,\,l\otimes l,\label{mp5}
\end{equation}
where $\rho^2=r^2+a_1^2\cos^2\theta+a_2^2\sin^2\theta$, and the null and geodesic vector $l$ is defined as 
\begin{equation}
l=dt+\frac{r^2\rho^2}{(r^2+a_1^2)(r^2+a_2^2)}dr-a_1\sin^2\theta d\varphi_1-a_2\cos^2\theta d\varphi_2.
\end{equation}
In order for our Buchdahl-Kerr-Schild composition to be valid in the presence of rotation, it is mandatory to turn off at least one of the angular momenta, let us say $a_2=0$ (the case $a_1=0$ can be treated identically).  Therefore, taking $\xi^{\mu}\partial_{\mu}=\partial_{\varphi_2}$ as the Killing vector of our transformation, our new spinning  vacuum solution ($R_{\mu\nu}=0$) in the Kerr-Schild form reads
\begin{align}
ds^2_{{{MPLC}}}=\left(r^2\cos^2\theta \right)\left[-dt^2+\frac{\rho_1^2}{r^2+a_1^2}dr^2+\rho_1^2d\theta^2+(r^2+a_1^2)\sin^2\theta d\varphi_1^2\right]+\frac{d\varphi_2^2}{r^2\cos^2\theta}-\frac{M r^2\cos^2\theta}{\rho_1^2}l_1\otimes l_1, \label{mp2}
\end{align}
where now $\rho_1^2=r^2+a_1^2\cos^2\theta$ and where the null and geodesic vector field $l_1$ is given by
\begin{equation}
l_1=dt+\frac{\rho_1^2}{(r^2+a_1^2)}dr-a_1\sin^2\theta d\varphi_1.
\end{equation}
Notice that the presence of a cosmic string localized at $\theta=\frac{\pi}{2}$ is unavoidable. Similarly, the construction could have been performed by turning off the angular momentum $a_1=0$, in such a case ending up with a cosmic string localized at $\theta=0$. In the more intuitive Boyer-Lindquist set of coordinates the line element \eqref{mp2} reads 
\begin{equation}
\begin{aligned}
ds^2_{MPLC}=\left(r^2\cos^2\theta \right)\left[-dt^2+\frac{M}{\rho_1^2}\left(dt-a_1\sin^2\theta d\varphi_1\right)^2+\rho_1^2\left(\frac{dr^2}{r^2+a_1^2-M}+d\theta^2\right) +(r^2+a_1^2)\sin^2\theta d\varphi_1^2\right]+\frac{d\varphi_2^2}{r^2\cos^2\theta}. \label{mp2BL}
\end{aligned}
\end{equation}
{The existence of the Boyer-Lindquist coordinates is a consequence of the circularity of the spacetime, the latter being a coordinate-independent property.}

A straightforward generalization of this result to the generic $d-$dimensional case proceeds directly by taking to zero one of the angular momenta. In the case of odd dimensions, $d=2n+1$, the Buchdahl-transformed Myers-Perry metric with $(n-1)$ rotations reads
\begin{equation}
\begin{aligned}
ds^2&=\left(r^2\mu_n^2\right)^{\frac{2}{d-3}}\Bigg[-dt^2+\frac{M r^2}{\Pi F}\left(dt+\sum_{i=1}^{n-1}a_i\mu_i^2d\varphi_i\right)^2+\frac{\Pi F}{\Pi-Mr^2}dr^2+\sum_{i=1}^{n-1}(r^2+a_i^2)\left(d\mu_i^2+\mu_i^2 d\varphi_i^2\right)+r^2 d\mu_n^2\Bigg]\\
&\quad+\frac{d\varphi_n^2}{r^2\mu_n^2},
\end{aligned}
\end{equation}
where 
\begin{equation}
\sum_{i=1}^n\mu_i^2=1,\qquad F=1-\sum_{i=1}^{n-1}\frac{a_i^2\mu_i^2}{r^2+a_i^2},\qquad \Pi=\prod_{i=1}^{n-1}(r^2+a_i^2)r^2.
\end{equation}
Similarly, for even dimensions $d=2n+2$ with $(n-1)$ rotations we obtain 
\begin{equation}
\begin{aligned}
ds^2&=\left(r^2\mu_n^2\right)^{\frac{2}{d-3}}\left[-dt^2+\frac{M r}{\Pi F}\left(dt+\sum_{i=1}^{n-1}a_i\mu_i^2d\varphi_i\right)^2+\frac{\Pi F}{\Pi-Mr}dr^2+\sum_{i=1}^{n-1}(r^2+a_i^2)\left(d\mu_i^2+\mu_i^2 d\varphi_i^2\right)+r^2 d\mu_n^2+r^2d\alpha^2\right]\\
&\quad+\frac{d\varphi_n^2}{r^2\mu_n^2},
\end{aligned}
\end{equation}
where the variables $(\mu_i, \alpha)$ are subject to the constraint $\alpha^2+\sum_{i=1}^n\mu_i^2=1$.

{It is interesting to remark that, 
it is natural to expect that these solutions are going to be of the general algebraic type according to the higher dimensional Petrov classification \cite{Coley:2004jv,Ortaggio:2012jd}. Again, the attractiveness of these solutions relies on the fact that they are vacuum algebraically general solutions admitting a Kerr-Schild decomposition. The Kerr-Schild ansatz can also be useful regarding what is called the double copy framework. This latter posits that scattering amplitudes in gauge theories, such as quantum chromodynamics, are intimately connected to those in gravity. On the other hand, while the Kerr-Schild ansatz is not directly applicable to the double copy in its usual formulation, there have been explorations of connections between exact solutions in gravity and their counterparts in gauge theories within the framework of the double copy. For example, certain properties of black hole solutions obtained through the Kerr-Schild ansatz may have analogs or implications in the context of the double copy (see e.g. \cite{Bahjat-Abbas:2017htu} and references therein and thereof).} This is precisely what we are going to investigate in the next subsection.

\subsection{Buchdahl-transformed Kerr-Schild metrics and double copy}

In the double copy scheme, the Kerr-Schild function $H$, employed for expressing the black hole metric as a perturbation of the seed metric along a null and geodesic direction $l$, becomes associated with the Maxwell gauge potential via $A\propto H l$. This relationship embodies the canonical correspondence inherent in the double copy formalism utilizing the Kerr-Schild ansatz. The underlying reasoning is to establish a framework for elucidating the interrelation between gauge theory amplitudes and gravitational counterparts, thereby facilitating the examination of scattering phenomena involving gravitons and their interactions with matter fields within specific spacetime configurations, notably those of rotating black holes.

In elucidating this correspondence, consider $g$ to represent a metric formulated within the Kerr-Schild paradigm
\begin{eqnarray}
g_{\mu\nu}=g^{(0)}_{\mu\nu}+H\ l_{\mu}l_{\nu},
\label{kmetric}
\end{eqnarray}
where $l$ is a null and geodesic vector field with respect to both metrics. In this case, the Ricci tensor transforms as
\begin{eqnarray}
R^{\mu}_{\,\nu}=R^{(0)\mu}_{\quad\,\,\nu}-H\,l^{\mu} l^{\sigma}R^{(0)}_{\sigma \nu}+\frac{1}{2}\nabla^{(0)}_{\sigma}\Big[\nabla^{(0)\mu}\left(H l^{\sigma}l_{\nu}\right)
+\nabla^{(0)}_{\nu}\left(H l^{\sigma}l^{\mu}\right)-\nabla^{(0)\sigma}\left(H l^{\mu}l_{\nu}\right)\Big].
\label{riccitransf}
\end{eqnarray}

Let us see what could be some hypotheses to ensure that the Kerr-Schild function $H$ of the Kerr-Schild metric (\ref{kmetric}) can generate a Maxwell solution with gauge potential $A\propto Hl$. We also remind that the null and geodesic vector fields can always be chosen such that $l_t=1$. For example, if the following conditions hold
\begin{eqnarray}
R_{\mu t}=R^{(0)}_{\mu t}=0,\qquad \Gamma^{(0)\mu}_{t\nu}=0,\qquad \partial_{t}g_{\mu\nu}=\partial_{t} H=\partial_{t} l_{\mu}=0,
\label{hypo}
\end{eqnarray}
the gauge potential $A\propto H\,l$ will satisfy the Maxwell equations. Indeed, taking the relation (\ref{riccitransf}) for $\nu=t$, that would imply that $\nabla^{(0)}_{t}\left(H l^{\sigma}l^{\mu}\right)=0$, and hence 
\begin{eqnarray}
\nabla^{(0)}_{\sigma}\Big[\nabla^{(0)\mu}\left(H l^{\sigma}\right)
-\nabla^{(0)\sigma}\left(H l^{\mu}\right)\Big]=\nabla^{(0)}_{\sigma}\,F^{\mu\sigma}=0.
\end{eqnarray}
It is a matter of checking to see that the sufficiency conditions (\ref{hypo}) hold in the case of the Schwarzschild metric  (\ref{schw})  and its Kerr extension with the ellipsoidal flat seed metric as well as for its higher-dimensional version  (\ref{mp5}). The legitimate question is what happens for the Schwarzschild-Levi-Civita (\ref{lv}) or the Buchdahl-transformed Myers-Perry solution  (\ref{mp2}).  A quick inspection reveals that in both cases, the condition $\Gamma^{(0)\mu}_{t\nu}=0$ is not fulfilled, and hence the sufficient conditions given in (\ref{hypo}) do not apply.  In fact, in the Schwarzschild-Levi-Civita black hole (\ref{lv}), the Kerr-Schild function and the null geodesic vector are $H_{{{{SLC}}}}(r,\theta)\propto r^3\,\sin^4\theta$ and $l=dt+dr$, respectively. 
Then one can  directly see that $A=H_{{{{SLC}}}}(r,\theta) (dt+dr)\propto  r^3\,\sin^4\theta  (dt+dr)$ will not satisfy the Maxwell equations for the Schwarzschild-Levi-Civita metric \eqref{lv}. But what is surprising is that by ignoring the term that generates the singularity along the axis, namely, the term $\sin^4\theta$, one can show that the gauge potential defined by
$$
{\cal A}=\frac{H_{{{{SLC}}}}(r,\theta)}{\sin^4\theta} (dt+dr)\propto  r^3\  (dt+dr),
$$
will be a solution of the Maxwell equations.  Our intuition is re-enforced with the example of the Myers-Perry solution transformed through the Buchdahl transformation (\ref{mp2}), where $H_{{{{MP}}}}(r,\theta)\propto \frac{r^2\cos^2\theta}{\rho_1^2}$ and the null geodesic vector is $l_1$. In this case again, one can see that the $1-$form gauge defined by 
$$
{\cal A}= \frac{H_{{{{MP}}}}(r,\theta)}{\cos^2\theta}\,l_1\propto \frac{r^2}{\rho_1^2}\left(dt+\frac{\rho_1^2}{(r^2+a_1^2)}dr-a_1\sin^2\theta d\varphi_1\right),
$$
will satisfy the Maxwell equations for the Myers-Perry Buchdahl-transformed metric (\ref{mp2}).

 We end this section by showing that although it is a modification of the simple copy gauge field $A\to {\cal A}$ that satisfies Maxwell's equations, the zeroth copy scalar field $H$ will satisfy itself the massless Klein-Gordon equation for the Buchdahl-transformed seed metric along the spacelike Killing vector field. Let us see how it works. For a vacuum Kerr-Schild metric of the form (\ref{kmetric}), it was shown \cite{Monteiro:2014cda} that the Kerr-Schild function $H$ would satisfy the massless Klein-Gordon equation for the seed metric $\Box_0 H=0$, where  $\Box_0$ means the d'Alembertian operator with respect to the seed metric. In our case, the Buchdahl-transformed Kerr-Schild metric is given by (\ref{ksm2}) with a seed metric given by
\begin{eqnarray}
\label{tg0}
d\tilde{s}_0^2=\left(g_{\varphi\varphi}^{(0)}\right)^{\frac{2}{d-3}} \left[g_{ij}^{(0)}dx^idx^j\right]+\frac{ d\varphi^2}{g_{\varphi\varphi}^{(0)}} .
\end{eqnarray}
It is now simple to see that since $\Box_0 H=0$ together with the fact that $\partial_{\varphi}H=0$, one also has that $\tilde{\Box}_0 H=0$, with $\tilde{\Box}_0$ being the d'Alembertian operator with respect to 
(\ref{tg0}). Indeed, this is a direct consequence of the fact that
$$
\sqrt{-\tilde{g}^{(0)}}=\sqrt{-{g}^{(0)}}\left(g_{\varphi\varphi}^{(0)}\right)^{\frac{2}{d-3}},\qquad \qquad\tilde{g}_{}^{(0)ij}=g_{}^{(0)ij} \left(g_{\varphi\varphi}^{(0)}\right)^{\frac{-2}{d-3}}.
$$

\section{Extending Buchdahl theorem in Einstein-Scalar theory} \label{sec4}

Let us now shift our focus to the Buchdahl theorem of the second kind \eqref{Buchdahl2}, which pertains to the Einstein-Scalar system in four dimensions \cite{Buchdahl:1959nk}. This theorem follows a similar line of argumentation as observed in the vacuum case.
Initially, it is noted that due to the specific characteristics of the scalar field, the transformation fails to dress Minkowski spacetime with a nontrivial scalar field. Conversely, by employing the Schwarzschild black hole as an initial configuration and employing (as usual) the transformation along the timelike Killing vector $\partial_t$, the theorem yields what is commonly referred to as the FJNW black hole \cite{Fisher:1948yn, Janis:1968zz, Janis:1969ivo}.\footnote{A three-parameter extension of the FJNW solution, including a rotating generalization, have been reported in \cite{Azizallahi:2023rrv, Mirza:2023mnm}. The static solution presented in \cite{Azizallahi:2023rrv} can be derived by the application of the Buchdahl theorem utilizing the Zipoy-Voorhees metric as a seed.} Various facets of this geometry have been studied \cite{Abdolrahimi:2009dc}, including its extension to higher dimensions within the Kaluza-Klein framework \cite{Lu:1995yn, Lu:1995sh, Bogush:2022qxl}. This, as elucidated in the subsequent section, furnishes an intriguing framework for applying a geometrical understanding of such solutions.

The FJNW spacetime is recognized for containing a curvature singularity precisely coinciding with the location of the would-be horizon at $r_h=2M$. Notably, irrespective of the parameter $\beta$ value, it yields Minkowski spacetime in the limit where the mass approaches zero. The presence of a naked singularity stems from the divergence of the associated energy-momentum tensor at $r_h=2M$, primarily attributed to the logarithmic behavior exhibited by the scalar field profile. However, it has been demonstrated that upon a frame transformation, specifically to a scalar theory characterized by a conformal coupling with the curvature, the divergence in the scalar field profile does not necessarily manifest as a divergence in the energy-momentum tensor. Consequently, the geometry remains free from naked singularities \cite{Bekenstein:1974sf}. For a specific choice of the hair parameter, $\beta=1/2$, this formulation reproduces the renowned BBMB solution \cite{Bocharova:1970skc,Bekenstein:1974sf}.

\subsection{The Fisher-Janis-Newman-Winicour-Levi-Civita spacetime}

As anticipated, investigations into the scalar-augmented version of the Buchdahl theorem have predominantly focused on scenarios where the transformation is applied along the timelike Killing vector $\partial_t$. Consequently, in this context, we proceed to formulate the hairy extension of the Schwarzschild-Levi-Civita black hole \cite{Mazharimousavi:2024hrg}. Its expression is given by\footnote{It is pertinent to note that the scalar field profile remains defined up to a sign. Consequently, this sign can be adjusted in accordance with the sign originating from its logarithmic behavior, thereby directing the term $r^2\sin^2\theta$ towards the numerator of the expression \eqref{HAIRYEXTENSION}.}
\begin{equation}\label{HAIRYEXTENSION}
\begin{aligned}
ds^2_{{{{FJNWLC}}}}&=\left(r^2\sin^2\theta\right)^{\beta+1}\left[-\left(1-\frac{2M}{r}\right)dt^2+\frac{dr^2}{1-\frac{2M}{r}}+r^2d\theta^2\right]+\frac{d\varphi^2}{\left(r^2\sin^2\theta\right)^\beta} ,\\
\Phi&=-\sqrt{\frac{1-\beta^2}{2\kappa}}\ln\left(r^2\sin^2\theta\right).
\end{aligned}
\end{equation}
This configuration serves as an extension of the FJNW solution \cite{Fisher:1948yn, Janis:1968zz, Janis:1969ivo}, derived through a Buchdahl transformation of the second kind applied to the spacelike Killing vector $\partial_\varphi$. Reaching the Schwarzschild-Levi-Civita black hole \eqref{lv2} in the limit $\beta=1$, we denote \eqref{HAIRYEXTENSION} as the FJNW-Levi-Civita spacetime. However, it is noteworthy that in the presence of a nontrivial scalar field profile, the metric \eqref{HAIRYEXTENSION} does not exhibit asymptotic behavior akin to that of a Levi-Civita spacetime, as there exists no value of $\beta$ yielding a Levi-Civita metric.
As previously noted, the Levi-Civita spacetime manifests a curvature singularity along the entirety of its symmetry axis, thereby inheriting the same characteristic in the Schwarzschild-Levi-Civita solution \eqref{lv2}. This same feature is preserved in the hairy extension \eqref{HAIRYEXTENSION}; an analysis of the Kretschmann scalar unveils
\begin{equation}
\begin{aligned}
R_{\mu\nu\rho\sigma}R^{\mu\nu\rho\sigma}&=\frac{1}{r^4\left(r\sin\theta\right)^{4+4\beta}}\left[4(\beta+1)\left(\frac{(\beta+1)(7\beta^2+2\beta+3)}{\sin^4\theta}-\frac{12M\beta(2\beta+1)}{r}\right)\right.\\
&\quad\left.-\frac{16M}{r}\left(\frac{(\beta+1)(7\beta^3+5\beta^2+3\beta+3)}{\sin^2\theta}-\frac{M(7\beta^4+20\beta^3+23\beta^2+10\beta+3)}{r}\right)\right],
\end{aligned}
\end{equation}
a divergent behavior for all possible values of $\beta$.
The scalar field profile exhibits regular behavior throughout the spacetime, except at the symmetry axis and at locations asymptotically distant from it. {While the standard FJNW solution presents a curvature singularity at $r_h=2M$, such a singularity is absent in this formulation.
Of significance is the observation that, asymptotically distant from the symmetry axis, the kinetic term of the scalar field remains finite, which, owing to the shift symmetry of the action, emerges as the pertinent quantity. Moreover, the energy density also retains finiteness in this regime. It is noteworthy that unlike the FJNW solution, wherein the massless limit yields Minkowski spacetime, the corresponding limit in the context of the FJNW-Levi-Civita solution results in a nontrivial geometry characterized by a nontrivial scalar field.}
Upon examination in cylindrical coordinates, it becomes apparent that this geometry cannot be reconciled with the Levi-Civita spacetime, as there exists no value of $\beta$ that sustains a nontrivial scalar field while simultaneously adhering to the Levi-Civita form for the geometry. The expression read
\begin{equation}
\begin{aligned}
    ds^2_{FJNWLC}&=\rho^{2(\beta+1)}(-dt^2+d\rho^2+dz^2)+\frac{d\varphi^2}{\rho^{2\beta}},\\
    \Phi&=-\sqrt{\frac{1-\beta^2}{2\kappa}}\ln\left(\rho^2\right).
\end{aligned}
\end{equation}
It turns out that the solution exhibits a special algebraic nature, classified as type D, thereby belonging to the broader Pleban\'ski-Demia\'nski class.  Conversely, for $\beta=-1/2$, it manifests as a conformally flat spacetime featuring a nontrivial scalar field, akin to the Levi-Civita spacetime with $\sigma=1$ if $\beta=1$. When $\beta=-1$, it simplifies to Minkowski spacetime.

An insightful approach to grasp the intricacies of this geometry involves its higher dimensional Kaluza-Klein uplift. Following the Kaluza-Klein framework (see e.g. \cite{Duff:1986hr}), whereby a solution of the Einstein-Scalar theory in dimension $d$ can always be extended to a higher dimensional counterpart pertaining to Einstein theory in dimension $d+1$, the uplifted geometry is given by
\begin{equation}
ds^2_{d+1}=e^{2\alpha\Phi}ds^2_d+e^{2\gamma\Phi}dz^2,
\end{equation}
being $z$ the extra spacetime coordinate and where 
\begin{equation}
    \alpha^2=\frac{1}{2(d-1)(d-2)},\qquad \gamma=-(d-2)\alpha.
\end{equation}
It is therefore natural to inquire what the higher dimensional extension of the FJNW-Levi-Civita solution \eqref{HAIRYEXTENSION} is and discern the potential geometrical insights stemming from such an extension. To embark upon this endeavor, let us commence with the FJNW-Levi-Civita solution expressed in cylindrical coordinates
\begin{equation}
\begin{aligned}
ds^2_{{{{FJNWLC}}}}&=\rho^{2(\beta+1)}ds^2_3+\frac{d\varphi^2}{\rho^{2\beta}},\\
\Phi&=-\sqrt{1-\beta^2}\ln(\rho^2),
\end{aligned}
\end{equation}
where we have used
\begin{equation}
    ds^2_3=-\frac{\sqrt{\rho^2+z^2}-2M}{\sqrt{\rho^2+z^2}} dt^2+\frac{\sqrt{\rho^2+z^2}}{\sqrt{\rho^2+z^2}-2M} \frac{(\rho d \rho+z d z)^2}{\rho^2+z^2}+\frac{(z d \rho-\rho d z)^2}{\rho^2+z^2}.
\end{equation}
To align with the conventions of \cite{Duff:1986hr}, we set $\kappa=1/2$. Consequently, the Kaluza-Klein five-dimensional extension of \eqref{HAIRYEXTENSION} is given by
\begin{equation}
ds_5^2=\rho^{2(\beta+1)-4\alpha\lambda}ds^2_3+\rho^{-4\alpha\lambda-2\beta}d\varphi^2+\frac{dz^2}{\rho^{4\gamma\lambda}},
\end{equation}
where we have introduced $\lambda=\sqrt{1-\beta^2}$, while in four dimensions, we set $\alpha=1/\sqrt{12}$ and $\gamma=-2/\sqrt{12}$.
This geometry reveals particularly insightful characteristics when opting for the hair parameter $\beta=-1/2$. Specifically, for this choice, we find that $2(\beta+1)-4\alpha\lambda=0=-4\alpha\lambda-2\beta$ and $4\gamma\lambda=-1$, thereby resulting in
\begin{equation}
    ds_5^2=-\left(1-\frac{2M}{r}\right)dt^2+\frac{dr^2}{\left(1-\frac{2M}{r}\right)}+r^2d\theta^2+r^2\sin^2\theta dz^2 + d\varphi^2,
\end{equation}
where for the sake of the physical interpretation of this line element we have turned back to the use of spherical coordinates. Consequently, it becomes evident that for $\beta=-1/2$, the Kaluza-Klein reduction from five to four dimensions of the standard homogeneous Schwarzschild black string, conducted along the angular direction $\varphi$, naturally yields the FJNW-Levi-Civita solution \eqref{HAIRYEXTENSION}. In simpler terms, the FJNW-Levi-Civita solution can be entirely regularized (for $\beta=-1/2$) through its uplift to five dimensions, where it manifests as the well-known homogeneous Schwarzschild black string. {It would be interesting to explore the stability of the dimensionally reduced configuration, given the fact that the five-dimensional uplift is unstable under long-wavelength perturbations.}

It is now natural to inquire whether a charged extension of this geometry can be devised following the framework of the JRW theorem \cite{Janis:1969ivo}. The JRW theorem offers a means to extend a nontrivial solution of the Einstein-Scalar system to incorporate electric charge, while leaving the scalar field profile unaffected by the electric field.
In direct parallel to this theorem, we introduce the following construction:
Let 
\begin{equation}\label{metricseed}
ds^2=e^{2U}h_{\alpha\beta}dx^\alpha dx^\beta+e^{-2U}d\varphi^2,
\end{equation}
be a four-dimensional metric that is ``static'' with respect to $\varphi$ and is a solution to the vacuum Einstein equations. Hence,
\begin{enumerate}[label=\textbf{(\roman*)}]
\item \label{enumerate1} The reciprocal metric
\begin{equation}
ds^2=e^{-2U}h_{\alpha\beta}dx^\alpha dx^\beta+e^{2U}d\varphi^2,
\end{equation}
is also a solution to the vacuum Einstein equations.
\item \label{enumerate2} If a vacuum solution to the Einstein equations is given by
\begin{equation}\label{metricscalar}
ds^2=e^{2V}h_{\alpha\beta}dx^\alpha dx^\beta+e^{-2V}d\varphi^2,
\end{equation}
then the metric \eqref{metricseed} is enhanced to be a solution to the Einstein-Scalar equations \eqref{ESeq}, where
\begin{equation}\label{scalarprop2}
\Phi=\sqrt{\frac{2}{\kappa}\frac{1-\beta^2}{\beta^2}}U,\qquad U=\beta V.
\end{equation}
\item \label{enumerate3} In addition, if a static solution to the Einstein-Scalar equations \eqref{ESeq} is given by \eqref{metricseed} and \eqref{scalarprop2}, then a solution to the Einstein-Scalar-Maxwell equations
\begin{equation}\label{einsteinscalarmaxwell}
R_{\mu\nu}=\kappa\left(\partial_\mu\Phi\partial_\nu\Phi+F_{\lambda\mu}F^{\lambda}_{\,\,\,\,\nu}-\frac{1}{4}g_{\mu\nu}F_{\lambda\rho}F^{\lambda\rho}\right),\quad
\square\Phi=0,\quad \nabla_\mu F^{\mu\nu}=0,
\end{equation}
is given by \eqref{scalarprop2} along with 
\begin{equation}
\begin{aligned}
ds^2&=e^{2W}h_{\alpha\beta}dx^\alpha dx^\beta+e^{-2W}d\varphi^2,\\
F_{\mu\nu}&=\sqrt{\frac{8}{\kappa}}e^{-2W}\left(U_{,\mu}\delta^\varphi_\nu-U_{,\nu}\delta^\varphi_\mu\right),
\end{aligned}
\end{equation}
where $W=\ln\left\vert2\cosh (U-U_0)\right\vert$, being $U_0$ a constant relevant for the determination of the nature of the magnetic configuration.\footnote{This has been noticed in \cite{Maeda:2019tqs} for the higher dimensional electric case, as it has been unnoticed in the original construction \cite{Janis:1969ivo}. Additionally, in \cite{Maeda:2016ddh} a detailed classification of higher-dimensional Einstein-Scalar-Maxwell solutions with Einstein base manifolds  has been provided.}
\end{enumerate}
Thus, from \ref{enumerate2} solution \eqref{HAIRYEXTENSION} is straightforwardly constructed, while from \ref{enumerate3} the magnetized extension of \eqref{HAIRYEXTENSION} is obtained, yielding 
\begin{equation}\label{MAGNETICHAIRYEXTENSION}
\begin{aligned}
ds^2&=\Lambda^2\left(r^2\sin^2\theta\right)^{\beta+1}\left[-\left(1-\frac{2M}{r}\right)dt^2+\frac{dr^2}{1-\frac{2M}{r}}+r^2d\theta^2\right]+\frac{1}{(r^2\sin^2\theta)^\beta\Lambda^2} d\varphi^2,\\
\Phi&=\sqrt{\frac{1-\beta^2}{2\kappa}}\ln\left(\frac{1}{r^2\sin^2\theta}\right),\quad A=\frac{B}{\Lambda}\left(\frac{1}{r^2\sin^2\theta}\right)^\beta d\varphi.\\
\end{aligned}
\end{equation}
Here it has been defined $\Lambda=1+B^2\left(\frac{1}{r^2\sin^2\theta}\right)^\beta$. 
The limit $\beta=1$ enables the recovery of the hairless magnetized Schwarzschild-Levi-Civita black hole \eqref{magneticbuchdahl}, while $\beta=-1$ yields the Schwarzschild-Melvin black hole.
In contrast to the approach in \cite{Janis:1969ivo}, our construction \ref{enumerate3} introduces a magnetic field rather than an electric one. {Specifically, it embeds the FJNW-Levi-Civita black hole within a Melvin universe, consistent with the ``magnetic" nature of the Levi-Civita black hole. However, it's important to note that this representation is essentially a concise depiction of the magnetized solution attainable through the application of magnetic Harrison transformations \cite{harrison1968new}, which are valid in the presence of a minimally coupled scalar field. While an electric Harrison transformation would also introduce a monopole electric charge, the compact construction \ref{enumerate3} would not be applicable. Instead, the approach of \cite{Janis:1969ivo} would be employed, which is formulated in an ``electric" context, as the Buchdahl transformations are applied with respect to the timelike Killing vector $\partial_t$.
Contrary to the hairless case \eqref{magneticbuchdahl}, the magnetic field does not regularize the axis of symmetry in the geometry \eqref{MAGNETICHAIRYEXTENSION}. The curvature singularity persists at the locus of the symmetry axis.}
\subsection{Conformal frame}

Up to this point, our focus has been on solutions within the Einstein-Scalar system. However, it is widely recognized that the Einstein-Scalar theory is conformally related to the Einstein-Conformal-Scalar theory. 
\begin{equation}
I_{ECS}=\int d^4x\sqrt{\bar{g}}\left[\frac{\bar{R}}{2\kappa}-\frac{1}{2}\left(\bar{\partial}\Psi\right)^2-\frac{1}{12}\bar{R}\Psi^2\right],
\end{equation}
via the so-called Bekenstein transformations \cite{Bekenstein:1974sf}
\begin{equation}\label{bekensteintrans}
\bar{g}_{\mu\nu}=\cosh^2\left(\sqrt{\frac{\kappa}{6}}\Phi\right)g_{\mu\nu},\quad \Psi=\sqrt{\frac{6}{\kappa}}\tanh\left(\sqrt{\frac{\kappa}{6}}\Phi\right).
\end{equation}
Hence, given a solution $(g_{\mu\nu},\Phi)$ of the Einstein-Scalar system, the set of transformations \eqref{bekensteintrans} promptly furnishes a solution $(\bar{g}_{\mu\nu},\Psi)$ for the improved set of field equations
\begin{equation}\label{eqsconformalframe}
\begin{aligned}
\bar{G}_{\mu\nu}&=\kappa\left[\bar{\partial}_\mu\Psi\bar{\partial}_\nu\Psi-\frac{1}{2}\bar{g}_{\mu\nu}\left(\bar{\partial}\Psi\right)^2+\frac{1}{6}\left(\bar{g}_{\mu\nu}\bar{\square}-\bar{\nabla}_\mu\bar{\nabla}_\nu+\bar{G}_{\mu\nu}\right)\Psi^2\right],\\
\bar{\square}\Psi&=\frac{1}{6}\bar{R}\Psi.
\end{aligned}
\end{equation}
The most iconic solution obtained through this approach is the renowned BBMB black hole \cite{Bocharova:1970skc,Bekenstein:1974sf}, which essentially corresponds to the FJNW solution translated to the conformal frame, particularly for the specific value $\beta=1/2$. Following this analogy, let us execute the change of frame on the FJNW-Levi-Civita spacetime \eqref{HAIRYEXTENSION}, focusing on the case of $\beta=1/2$. This yields 
\begin{equation}\label{solconformalframe}
\begin{aligned}
d\bar{s}^2&=\left(1+\alpha r\sin\theta\right)^2\left[r^2\sin^2\theta\left[-\left(1-\frac{2M}{r}\right)dt^2+\frac{dr^2}{\left(1-\frac{2M}{r}\right)}+r^2d\theta^2\right]+\frac{d\varphi^2}{r^2\sin^2\theta}\right],\\
\Psi&=\sqrt{\frac{6}{\kappa}}\frac{1-\alpha r\sin\theta}{1+\alpha r\sin\theta},
\end{aligned}
\end{equation}
where $\alpha>0$ is a constant that can be introduced after the rescalings $(t,r,\varphi)\rightarrow (\alpha t,\alpha r,\alpha^3\varphi)$ together with $M\rightarrow \alpha M$.
{The solution turns out to be of algebraic type I, retaining a curvature singularity at the axis of symmetry, a characteristic inherent in the solutions we have considered so far. However, the axis of symmetry can be regularized by immersing the solution into an exterior magnetic field, see below \eqref{LCBBMBMELVIN}. At distances asymptotically distant from the symmetry axis, both the kinetic term and energy density remain finite. Conversely, the scalar field has been entirely regularized, exhibiting regular behavior across the entirety of the domain of outer communications, including the symmetry axis.}

An intriguing scenario unfolds when $M=0$. In this instance, the spacetime diverges from the Minkowski geometry yet retains a nontrivial scalar field, deviating from the behavior demonstrated by the BBMB solution. The solution in this case can be represented as
\begin{equation}
\begin{aligned}
d\bar{s}^2&=\left(1+\alpha r\sin\theta\right)^2\left[r^2\sin^2\theta\left(-dt^2+dr^2+r^2d\theta^2\right)+\frac{d\varphi^2}{r^2\sin^2\theta}\right],\\
\Psi&=\sqrt{\frac{6}{\kappa}}\frac{1-\alpha r\sin\theta}{1+\alpha r\sin\theta}.
\end{aligned}
\end{equation}
{This solution also features a singularity along the symmetry axis; however, it falls into type D in the algebraic classification, unlike \eqref{solconformalframe}. Although the introduction of the parameter $\alpha$ may seem somewhat arbitrary, it proves instrumental in analyzing the spacetime emerging in the limit as $\alpha$ tends to zero.
The solution maintains its type I classification in the Petrov scheme.} Notably, it does not straightforwardly connect with either the Schwarzschild or the Levi-Civita black hole, partly due to the constant scalar field present
\begin{equation}
\begin{aligned}
d\bar{s}^2&=r^2\sin^2\theta\left[-\left(1-\frac{2M}{r}\right)dt^2+\frac{dr^2}{\left(1-\frac{2M}{r}\right)}+r^2d\theta^2\right]+\frac{d\varphi^2}{r^2\sin^2\theta},\\
\Psi&=\sqrt{\frac{6}{\kappa}}.
\end{aligned}
\end{equation}
The scalar field constant value is precisely the one that makes the effective gravitational constant vanish. Therefore, this solution solves the field equations \eqref{eqsconformalframe} in the limit where no Einstein term is present in the action.

As previously mentioned, \eqref{solconformalframe} can be regularized along its symmetry axis by embedding the solution into an external magnetic field. The most direct method to achieve this is by considering the already magnetized solution \eqref{MAGNETICHAIRYEXTENSION} and applying the change of frame provided by \eqref{bekensteintrans}. 
The solution that solves the Einstein-Conformal-Scalar-Maxwell equations then reads
\begin{equation}
\begin{aligned}\label{LCBBMBMELVIN}
d\bar{s}^2&=\left(1+\alpha r\sin\theta\right)^2\left[\Lambda^2 \left[-\left(1-\frac{2M}{r}\right)dt^2+\frac{dr^2}{\left(1-\frac{2M}{r}\right)}+r^2d\theta^2\right]+\frac{d\varphi^2}{\Lambda^2}\right],\\
\Psi&=\sqrt{\frac{6}{\kappa}}\frac{1-\alpha r\sin\theta}{1+\alpha r\sin\theta},\quad A=\frac{2\sqrt{\alpha}B}{\Lambda}d\varphi,
\end{aligned}
\end{equation}
where $\Lambda=r\sin\theta+B^2$. {The inclusion of the magnetic field regularizes the symmetry axis and the curvature singularity locates itself at $r=0$ only. Hence, this geometry remains regular everywhere within the domain of outer communications, as well as the scalar field profile. This stands in contrast to the standard BBMB magnetized black hole.}

\section{Further Comments}\label{sec5}

In this work, we have revisited Buchdahl transformations  \cite{Buchdahl:1956zz,Buchdahl:1959nk}, specifically exploring scenarios where the transformations are applied with respect to the typical azimuthal spacelike Killing vector, as opposed to the customary timelike Killing vector. {This has allowed us to identify an infinite dimensional, numerable, non-Abelian subgroup of the group of hidden symmetries of the Einstein equations for stationary, axisymmetric spacetimes.} This seemingly minor adjustment has paved the way for the construction and examination of novel black hole solutions in both vacuum and the Einstein-Scalar system, as well as its conformally related framework. Indeed, we have elucidated the origin of the recently introduced spacetime \cite{Mazharimousavi:2024hrg} and delved into Buchdahl transformations of the first kind \cite{Buchdahl:1956zz} in cases where the seed geometry possesses two different Killing vectors. Additionally, we have composed Buchdahl and Kerr-Schild transformations to unveil the Kerr-Schild form of these newly discovered algebraically general geometries, which, to the best of our knowledge, constitute the first vacuum Kerr-Schild geometries of type I.

This composition has provided valuable insights into the construction of new vacuum rotating black holes, particularly exemplified by what we have termed the Levi-Civita extension of the Myers-Perry geometry \cite{Myers:1986un}. Moreover, we have identified an algebraically general double copy framework.

Within the realm of Buchdahl transformations of the second kind \cite{Buchdahl:1959nk}, we have devised novel black hole solutions in the Einstein-Scalar and Einstein-Scalar-Maxwell theories. Notably, we have introduced what we refer to as a Levi-Civita extension of the FJNW solution \cite{Fisher:1948yn,Janis:1968zz}, or alternatively, a hairy extension of the Schwarzschild-Levi-Civita black hole \cite{Mazharimousavi:2024hrg}. Furthermore, we have extended the JRW theorem \cite{Janis:1969ivo} to embed this geometry within an external magnetic field. Finally, these Einstein-Scalar solutions have been uplifted to the Einstein-Conformal-Scalar framework, yielding a generalization of the well-known BBMB geometry \cite{Bocharova:1970skc,Bekenstein:1974sf}.

Numerous avenues of exploration emerge from these findings. The study of spacetime with acceleration \cite{Kinnersley:1970zw,Pravda:2000vh} proves to be particularly promising, offering a highly nontrivial static seed spacetime for investigation in this context. This could potentially lead to a C-metric spacetime embedded within a Levi-Civita-like background, possibly altering the form of the standard conical singularities presented in the metric. Since these are vacuum solutions, Harrison transformations \cite{harrison1968new} can be readily applied, facilitating the straightforward construction of their charged extensions with both electric and magnetic monopolic charges. Furthermore, it would be feasible to embed them within an electromagnetic background \cite{Bonnor_1954,Melvin1966}. Similarly, employing Ehlers transformations \cite{Ehlers:1957zz,Ehlers:1959aug}, these solutions could acquire properties of stationarity, such as NUT charge or vortex-like background characteristics \cite{Barrientos:2024pkt}. Investigating the causal structures and geometric features of these spacetimes would undoubtedly represent a significant advancement in determining the spectrum of solutions of Einstein vacuum gravity.
Transitioning to scenarios involving rotation, a comprehensive geometrical analysis of the Levi-Civita extension of the Myers-Perry black hole appears imperative. Moreover, given that we have access to Buchdahl transformations of the second kind and that the use of Ehlers and Harrison transformations is ensured within the Einstein-Scalar framework, all aforementioned constructions can be executed in the presence of a massless minimally coupled scalar field. Consequently, within the realm of very minimalistic scalar-tensor theories, the spectrum of exact solutions seems notably enriched by these discoveries, particularly when considering the utilization of frame transformations \cite{Bekenstein:1974sf,Ayon-Beato:2015ada}.
A subsequent step could involve augmenting the theory by including a cosmological constant, as demonstrated in \cite{Tangen:2007yn}. With this extension, not only can solutions of the type considered and mentioned here be utilized as seeds, but also cylindrically symmetric black holes \cite{Lemos:1994xp}. It would then be feasible to apply transformations along the translational symmetry Killing vector $\partial_z$, either independently or in combination with the usual Killing vectors $\partial_t$ and $\partial_\varphi$.

We anticipate that exploring these additional avenues of research will provide us with a more comprehensive understanding of the spectrum of exact solutions in Einstein and Einstein-Scalar theories.

\acknowledgments
We thank Andr\'es Anabal\'on for interesting discussions regarding the Kaluza-Klein uplift of \eqref{HAIRYEXTENSION}. The work of J.B. is supported by FONDECYT Postdoctorado grant 3230596. A.C. is partially supported by FONDECYT grant 1210500 and by PRIMUS/23/SCI/005 and GAČR 22-14791S grants from Charles University.  The work of M.H. is partially supported by FONDECYT grant 1210889. J.O. is partially supported by FONDECYT grant 1221504.

\nocite{*}
\bibliography{apssamp}

\begin{thebibliography}{44}
\expandafter\ifx\csname natexlab\endcsname\relax\def\natexlab#1{#1}\fi
\expandafter\ifx\csname bibnamefont\endcsname\relax
  \def\bibnamefont#1{#1}\fi
\expandafter\ifx\csname bibfnamefont\endcsname\relax
  \def\bibfnamefont#1{#1}\fi
\expandafter\ifx\csname citenamefont\endcsname\relax
  \def\citenamefont#1{#1}\fi
\expandafter\ifx\csname url\endcsname\relax
  \def\url#1{\texttt{#1}}\fi
\expandafter\ifx\csname urlprefix\endcsname\relax\def\urlprefix{URL }\fi
\providecommand{\bibinfo}[2]{#2}
\providecommand{\eprint}[2][]{\url{#2}}

\bibitem[{\citenamefont{Stephani et~al.}(2003)\citenamefont{Stephani, Kramer, MacCallum, Hoenselaers, and Herlt}}]{Stephani:2003tm}
\bibinfo{author}{\bibfnamefont{H.}~\bibnamefont{Stephani}}, \bibinfo{author}{\bibfnamefont{D.}~\bibnamefont{Kramer}}, \bibinfo{author}{\bibfnamefont{M.~A.~H.} \bibnamefont{MacCallum}}, \bibinfo{author}{\bibfnamefont{C.}~\bibnamefont{Hoenselaers}}, \bibnamefont{and} \bibinfo{author}{\bibfnamefont{E.}~\bibnamefont{Herlt}}, \emph{\bibinfo{title}{{Exact solutions of Einstein's field equations}}}, Cambridge Monographs on Mathematical Physics (\bibinfo{publisher}{Cambridge Univ. Press}, \bibinfo{address}{Cambridge}, \bibinfo{year}{2003}).

\bibitem[{\citenamefont{Ernst}(1968{\natexlab{a}})}]{Ernst:1967wx}
\bibinfo{author}{\bibfnamefont{F.~J.} \bibnamefont{Ernst}}, \bibinfo{journal}{Phys. Rev.} \textbf{\bibinfo{volume}{167}}, \bibinfo{pages}{1175} (\bibinfo{year}{1968}{\natexlab{a}}).

\bibitem[{\citenamefont{Ernst}(1968{\natexlab{b}})}]{Ernst:1967by}
\bibinfo{author}{\bibfnamefont{F.~J.} \bibnamefont{Ernst}}, \bibinfo{journal}{Phys. Rev.} \textbf{\bibinfo{volume}{168}}, \bibinfo{pages}{1415} (\bibinfo{year}{1968}{\natexlab{b}}).

\bibitem[{\citenamefont{Geroch}(1971)}]{Geroch:1970nt}
\bibinfo{author}{\bibfnamefont{R.~P.} \bibnamefont{Geroch}}, \bibinfo{journal}{J. Math. Phys.} \textbf{\bibinfo{volume}{12}}, \bibinfo{pages}{918} (\bibinfo{year}{1971}).

\bibitem[{\citenamefont{{Kinnersley}}(1973)}]{Kinn1}
\bibinfo{author}{\bibfnamefont{W.}~\bibnamefont{{Kinnersley}}}, \bibinfo{journal}{J. Math. Phys.} \textbf{\bibinfo{volume}{14}}, \bibinfo{pages}{651} (\bibinfo{year}{1973}).

\bibitem[{\citenamefont{Ehlers}(1957)}]{Ehlers:1957zz}
\bibinfo{author}{\bibfnamefont{J.}~\bibnamefont{Ehlers}}, \bibinfo{type}{Other thesis} (\bibinfo{year}{1957}).

\bibitem[{\citenamefont{Ehlers}(1962)}]{Ehlers:1959aug}
\bibinfo{author}{\bibfnamefont{J.}~\bibnamefont{Ehlers}}, \bibinfo{journal}{Colloq. Int. CNRS} \textbf{\bibinfo{volume}{91}}, \bibinfo{pages}{275} (\bibinfo{year}{1962}).

\bibitem[{\citenamefont{Harrison}(1968)}]{harrison1968new}
\bibinfo{author}{\bibfnamefont{B.~K.} \bibnamefont{Harrison}}, \bibinfo{journal}{J. Math. Phys.} \textbf{\bibinfo{volume}{9}}, \bibinfo{pages}{1744} (\bibinfo{year}{1968}).

\bibitem[{\citenamefont{Buchdahl}(1956)}]{Buchdahl:1956zz}
\bibinfo{author}{\bibfnamefont{H.~A.} \bibnamefont{Buchdahl}}, \bibinfo{journal}{Austral. J. Phys.} \textbf{\bibinfo{volume}{9}}, \bibinfo{pages}{13} (\bibinfo{year}{1956}).

\bibitem[{\citenamefont{Buchdahl}(1959)}]{Buchdahl:1959nk}
\bibinfo{author}{\bibfnamefont{H.~A.} \bibnamefont{Buchdahl}}, \bibinfo{journal}{Phys. Rev.} \textbf{\bibinfo{volume}{115}}, \bibinfo{pages}{1325} (\bibinfo{year}{1959}).

\bibitem[{\citenamefont{Janis et~al.}(1969)\citenamefont{Janis, Robinson, and Winicour}}]{Janis:1969ivo}
\bibinfo{author}{\bibfnamefont{A.~I.} \bibnamefont{Janis}}, \bibinfo{author}{\bibfnamefont{D.~C.} \bibnamefont{Robinson}}, \bibnamefont{and} \bibinfo{author}{\bibfnamefont{J.}~\bibnamefont{Winicour}}, \bibinfo{journal}{Phys. Rev.} \textbf{\bibinfo{volume}{186}}, \bibinfo{pages}{1729} (\bibinfo{year}{1969}).

\bibitem[{\citenamefont{Fisher}(1948)}]{Fisher:1948yn}
\bibinfo{author}{\bibfnamefont{I.~Z.} \bibnamefont{Fisher}}, \bibinfo{journal}{Zh. Eksp. Teor. Fiz.} \textbf{\bibinfo{volume}{18}}, \bibinfo{pages}{636} (\bibinfo{year}{1948}), \eprint{gr-qc/9911008}.

\bibitem[{\citenamefont{Janis et~al.}(1968)\citenamefont{Janis, Newman, and Winicour}}]{Janis:1968zz}
\bibinfo{author}{\bibfnamefont{A.~I.} \bibnamefont{Janis}}, \bibinfo{author}{\bibfnamefont{E.~T.} \bibnamefont{Newman}}, \bibnamefont{and} \bibinfo{author}{\bibfnamefont{J.}~\bibnamefont{Winicour}}, \bibinfo{journal}{Phys. Rev. Lett.} \textbf{\bibinfo{volume}{20}}, \bibinfo{pages}{878} (\bibinfo{year}{1968}).

\bibitem[{\citenamefont{Mazharimousavi}(2024)}]{Mazharimousavi:2024hrg}
\bibinfo{author}{\bibfnamefont{S.~H.} \bibnamefont{Mazharimousavi}} (\bibinfo{year}{2024}), \eprint{2403.02365}.

\bibitem[{\citenamefont{Plebanski and Demianski}(1976)}]{Plebanski:1976gy}
\bibinfo{author}{\bibfnamefont{J.~F.} \bibnamefont{Plebanski}} \bibnamefont{and} \bibinfo{author}{\bibfnamefont{M.}~\bibnamefont{Demianski}}, \bibinfo{journal}{Annals Phys.} \textbf{\bibinfo{volume}{98}}, \bibinfo{pages}{98} (\bibinfo{year}{1976}).

\bibitem[{\citenamefont{Goldberg and Sachs}(2009)}]{GS}
\bibinfo{author}{\bibfnamefont{J.~N.} \bibnamefont{Goldberg}} \bibnamefont{and} \bibinfo{author}{\bibfnamefont{R.~K.} \bibnamefont{Sachs}}, \bibinfo{journal}{Gen. Relativ. Gravit} \textbf{\bibinfo{volume}{41}}, \bibinfo{pages}{433–444} (\bibinfo{year}{2009}).

\bibitem[{\citenamefont{Myers and Perry}(1986)}]{Myers:1986un}
\bibinfo{author}{\bibfnamefont{R.~C.} \bibnamefont{Myers}} \bibnamefont{and} \bibinfo{author}{\bibfnamefont{M.~J.} \bibnamefont{Perry}}, \bibinfo{journal}{Annals Phys.} \textbf{\bibinfo{volume}{172}}, \bibinfo{pages}{304} (\bibinfo{year}{1986}).

\bibitem[{\citenamefont{Bern et~al.}(2008)\citenamefont{Bern, Carrasco, and Johansson}}]{Bern:2008qj}
\bibinfo{author}{\bibfnamefont{Z.}~\bibnamefont{Bern}}, \bibinfo{author}{\bibfnamefont{J.~J.~M.} \bibnamefont{Carrasco}}, \bibnamefont{and} \bibinfo{author}{\bibfnamefont{H.}~\bibnamefont{Johansson}}, \bibinfo{journal}{Phys. Rev. D} \textbf{\bibinfo{volume}{78}}, \bibinfo{pages}{085011} (\bibinfo{year}{2008}), \eprint{0805.3993}.

\bibitem[{\citenamefont{Bern et~al.}(2010)\citenamefont{Bern, Carrasco, and Johansson}}]{Bern:2010ue}
\bibinfo{author}{\bibfnamefont{Z.}~\bibnamefont{Bern}}, \bibinfo{author}{\bibfnamefont{J.~J.~M.} \bibnamefont{Carrasco}}, \bibnamefont{and} \bibinfo{author}{\bibfnamefont{H.}~\bibnamefont{Johansson}}, \bibinfo{journal}{Phys. Rev. Lett.} \textbf{\bibinfo{volume}{105}}, \bibinfo{pages}{061602} (\bibinfo{year}{2010}), \eprint{1004.0476}.

\bibitem[{\citenamefont{Monteiro et~al.}(2014)\citenamefont{Monteiro, O'Connell, and White}}]{Monteiro:2014cda}
\bibinfo{author}{\bibfnamefont{R.}~\bibnamefont{Monteiro}}, \bibinfo{author}{\bibfnamefont{D.}~\bibnamefont{O'Connell}}, \bibnamefont{and} \bibinfo{author}{\bibfnamefont{C.~D.} \bibnamefont{White}}, \bibinfo{journal}{JHEP} \textbf{\bibinfo{volume}{12}}, \bibinfo{pages}{056} (\bibinfo{year}{2014}), \eprint{1410.0239}.

\bibitem[{\citenamefont{Bekenstein}(1974)}]{Bekenstein:1974sf}
\bibinfo{author}{\bibfnamefont{J.~D.} \bibnamefont{Bekenstein}}, \bibinfo{journal}{Annals Phys.} \textbf{\bibinfo{volume}{82}}, \bibinfo{pages}{535} (\bibinfo{year}{1974}).

\bibitem[{\citenamefont{Bocharova et~al.}(1970)\citenamefont{Bocharova, Bronnikov, and Melnikov}}]{Bocharova:1970skc}
\bibinfo{author}{\bibfnamefont{N.~M.} \bibnamefont{Bocharova}}, \bibinfo{author}{\bibfnamefont{K.~A.} \bibnamefont{Bronnikov}}, \bibnamefont{and} \bibinfo{author}{\bibfnamefont{V.~N.} \bibnamefont{Melnikov}}, \bibinfo{journal}{.Vestn.Mosk.Univ.Ser.III Fiz.Astron.} \textbf{\bibinfo{volume}{6}}, \bibinfo{pages}{706} (\bibinfo{year}{1970}).

\bibitem[{\citenamefont{Levi-Civita}(1919)}]{LC}
\bibinfo{author}{\bibfnamefont{T.}~\bibnamefont{Levi-Civita}}, \bibinfo{journal}{Rend. Accad. Lincei} \textbf{\bibinfo{volume}{28}}, \bibinfo{pages}{101–109. §§7.3, 10.2} (\bibinfo{year}{1919}).

\bibitem[{\citenamefont{Bonnor}(1954)}]{Bonnor_1954}
\bibinfo{author}{\bibfnamefont{W.~B.} \bibnamefont{Bonnor}}, \bibinfo{journal}{Proceedings of the Physical Society. Section A} \textbf{\bibinfo{volume}{67}}, \bibinfo{pages}{225} (\bibinfo{year}{1954}).

\bibitem[{\citenamefont{Melvin and Wallingford}(1966)}]{Melvin1966}
\bibinfo{author}{\bibfnamefont{M.~A.} \bibnamefont{Melvin}} \bibnamefont{and} \bibinfo{author}{\bibfnamefont{J.~S.} \bibnamefont{Wallingford}}, \bibinfo{journal}{J. Math. Phys.} \textbf{\bibinfo{volume}{7}}, \bibinfo{pages}{333} (\bibinfo{year}{1966}).

\bibitem[{\citenamefont{Hassa\"\i{}ne}(2015)}]{Hassaine:2015ifa}
\bibinfo{author}{\bibfnamefont{M.}~\bibnamefont{Hassa\"\i{}ne}}, \bibinfo{journal}{Phys. Rev. D} \textbf{\bibinfo{volume}{91}}, \bibinfo{pages}{084054} (\bibinfo{year}{2015}), \eprint{1503.01716}.

\bibitem[{\citenamefont{Coley et~al.}(2004)\citenamefont{Coley, Milson, Pravda, and Pravdova}}]{Coley:2004jv}
\bibinfo{author}{\bibfnamefont{A.}~\bibnamefont{Coley}}, \bibinfo{author}{\bibfnamefont{R.}~\bibnamefont{Milson}}, \bibinfo{author}{\bibfnamefont{V.}~\bibnamefont{Pravda}}, \bibnamefont{and} \bibinfo{author}{\bibfnamefont{A.}~\bibnamefont{Pravdova}}, \bibinfo{journal}{Class. Quant. Grav.} \textbf{\bibinfo{volume}{21}}, \bibinfo{pages}{L35} (\bibinfo{year}{2004}), \eprint{gr-qc/0401008}.

\bibitem[{\citenamefont{Ortaggio et~al.}(2013)\citenamefont{Ortaggio, Pravda, and Pravdova}}]{Ortaggio:2012jd}
\bibinfo{author}{\bibfnamefont{M.}~\bibnamefont{Ortaggio}}, \bibinfo{author}{\bibfnamefont{V.}~\bibnamefont{Pravda}}, \bibnamefont{and} \bibinfo{author}{\bibfnamefont{A.}~\bibnamefont{Pravdova}}, \bibinfo{journal}{Class. Quant. Grav.} \textbf{\bibinfo{volume}{30}}, \bibinfo{pages}{013001} (\bibinfo{year}{2013}), \eprint{1211.7289}.

\bibitem[{\citenamefont{Bahjat-Abbas et~al.}(2017)\citenamefont{Bahjat-Abbas, Luna, and White}}]{Bahjat-Abbas:2017htu}
\bibinfo{author}{\bibfnamefont{N.}~\bibnamefont{Bahjat-Abbas}}, \bibinfo{author}{\bibfnamefont{A.}~\bibnamefont{Luna}}, \bibnamefont{and} \bibinfo{author}{\bibfnamefont{C.~D.} \bibnamefont{White}}, \bibinfo{journal}{JHEP} \textbf{\bibinfo{volume}{12}}, \bibinfo{pages}{004} (\bibinfo{year}{2017}), \eprint{1710.01953}.

\bibitem[{\citenamefont{Azizallahi et~al.}(2024)\citenamefont{Azizallahi, Mirza, Hajibarat, and Anjomshoa}}]{Azizallahi:2023rrv}
\bibinfo{author}{\bibfnamefont{A.}~\bibnamefont{Azizallahi}}, \bibinfo{author}{\bibfnamefont{B.}~\bibnamefont{Mirza}}, \bibinfo{author}{\bibfnamefont{A.}~\bibnamefont{Hajibarat}}, \bibnamefont{and} \bibinfo{author}{\bibfnamefont{H.}~\bibnamefont{Anjomshoa}}, \bibinfo{journal}{Nucl. Phys. B} \textbf{\bibinfo{volume}{998}}, \bibinfo{pages}{116414} (\bibinfo{year}{2024}), \eprint{2307.09328}.

\bibitem[{\citenamefont{Mirza et~al.}(2023)\citenamefont{Mirza, Kangazi, and Sadeghi}}]{Mirza:2023mnm}
\bibinfo{author}{\bibfnamefont{B.}~\bibnamefont{Mirza}}, \bibinfo{author}{\bibfnamefont{P.~K.} \bibnamefont{Kangazi}}, \bibnamefont{and} \bibinfo{author}{\bibfnamefont{F.}~\bibnamefont{Sadeghi}}, \bibinfo{journal}{Eur. Phys. J. C} \textbf{\bibinfo{volume}{83}}, \bibinfo{pages}{1161} (\bibinfo{year}{2023}), \eprint{2307.13588}.

\bibitem[{\citenamefont{Abdolrahimi and Shoom}(2010)}]{Abdolrahimi:2009dc}
\bibinfo{author}{\bibfnamefont{S.}~\bibnamefont{Abdolrahimi}} \bibnamefont{and} \bibinfo{author}{\bibfnamefont{A.~A.} \bibnamefont{Shoom}}, \bibinfo{journal}{Phys. Rev. D} \textbf{\bibinfo{volume}{81}}, \bibinfo{pages}{024035} (\bibinfo{year}{2010}), \eprint{0911.5380}.

\bibitem[{\citenamefont{Lu and Pope}(1996)}]{Lu:1995yn}
\bibinfo{author}{\bibfnamefont{H.}~\bibnamefont{Lu}} \bibnamefont{and} \bibinfo{author}{\bibfnamefont{C.~N.} \bibnamefont{Pope}}, \bibinfo{journal}{Nucl. Phys. B} \textbf{\bibinfo{volume}{465}}, \bibinfo{pages}{127} (\bibinfo{year}{1996}), \eprint{hep-th/9512012}.

\bibitem[{\citenamefont{Lu and Pope}(1997)}]{Lu:1995sh}
\bibinfo{author}{\bibfnamefont{H.}~\bibnamefont{Lu}} \bibnamefont{and} \bibinfo{author}{\bibfnamefont{C.~N.} \bibnamefont{Pope}}, \bibinfo{journal}{Int. J. Mod. Phys. A} \textbf{\bibinfo{volume}{12}}, \bibinfo{pages}{437} (\bibinfo{year}{1997}), \eprint{hep-th/9512153}.

\bibitem[{\citenamefont{Bogush and Gal'tsov}(2022)}]{Bogush:2022qxl}
\bibinfo{author}{\bibfnamefont{I.}~\bibnamefont{Bogush}} \bibnamefont{and} \bibinfo{author}{\bibfnamefont{D.}~\bibnamefont{Gal'tsov}}, \bibinfo{journal}{Phys. Rev. D} \textbf{\bibinfo{volume}{106}}, \bibinfo{pages}{084054} (\bibinfo{year}{2022}), \eprint{2208.14667}.

\bibitem[{\citenamefont{Duff et~al.}(1986)\citenamefont{Duff, Nilsson, and Pope}}]{Duff:1986hr}
\bibinfo{author}{\bibfnamefont{M.~J.} \bibnamefont{Duff}}, \bibinfo{author}{\bibfnamefont{B.~E.~W.} \bibnamefont{Nilsson}}, \bibnamefont{and} \bibinfo{author}{\bibfnamefont{C.~N.} \bibnamefont{Pope}}, \bibinfo{journal}{Phys. Rept.} \textbf{\bibinfo{volume}{130}}, \bibinfo{pages}{1} (\bibinfo{year}{1986}).

\bibitem[{\citenamefont{Maeda and Martinez}(2019)}]{Maeda:2019tqs}
\bibinfo{author}{\bibfnamefont{H.}~\bibnamefont{Maeda}} \bibnamefont{and} \bibinfo{author}{\bibfnamefont{C.}~\bibnamefont{Martinez}}, \bibinfo{journal}{Class. Quant. Grav.} \textbf{\bibinfo{volume}{36}}, \bibinfo{pages}{185017} (\bibinfo{year}{2019}), \eprint{1904.01658}.

\bibitem[{\citenamefont{Maeda and Martinez}(2018)}]{Maeda:2016ddh}
\bibinfo{author}{\bibfnamefont{H.}~\bibnamefont{Maeda}} \bibnamefont{and} \bibinfo{author}{\bibfnamefont{C.}~\bibnamefont{Martinez}}, \bibinfo{journal}{Eur. Phys. J. C} \textbf{\bibinfo{volume}{78}}, \bibinfo{pages}{860} (\bibinfo{year}{2018}), \eprint{1603.03436}.

\bibitem[{\citenamefont{Kinnersley and Walker}(1970)}]{Kinnersley:1970zw}
\bibinfo{author}{\bibfnamefont{W.}~\bibnamefont{Kinnersley}} \bibnamefont{and} \bibinfo{author}{\bibfnamefont{M.}~\bibnamefont{Walker}}, \bibinfo{journal}{Phys. Rev. D} \textbf{\bibinfo{volume}{2}}, \bibinfo{pages}{1359} (\bibinfo{year}{1970}).

\bibitem[{\citenamefont{Pravda and Pravdova}(2000)}]{Pravda:2000vh}
\bibinfo{author}{\bibfnamefont{V.}~\bibnamefont{Pravda}} \bibnamefont{and} \bibinfo{author}{\bibfnamefont{A.}~\bibnamefont{Pravdova}}, \bibinfo{journal}{Czech. J. Phys.} \textbf{\bibinfo{volume}{50}}, \bibinfo{pages}{333} (\bibinfo{year}{2000}), \eprint{gr-qc/0003067}.

\bibitem[{\citenamefont{Barrientos et~al.}(2024)\citenamefont{Barrientos, Cisterna, Kol\'a\v{r}, M\"uller, Oyarzo, and Pallikaris}}]{Barrientos:2024pkt}
\bibinfo{author}{\bibfnamefont{J.}~\bibnamefont{Barrientos}}, \bibinfo{author}{\bibfnamefont{A.}~\bibnamefont{Cisterna}}, \bibinfo{author}{\bibfnamefont{I.}~\bibnamefont{Kol\'a\v{r}}}, \bibinfo{author}{\bibfnamefont{K.}~\bibnamefont{M\"uller}}, \bibinfo{author}{\bibfnamefont{M.}~\bibnamefont{Oyarzo}}, \bibnamefont{and} \bibinfo{author}{\bibfnamefont{K.}~\bibnamefont{Pallikaris}}, \bibinfo{journal}{Eur. Phys. J. C} \textbf{\bibinfo{volume}{84}}, \bibinfo{pages}{724} (\bibinfo{year}{2024}), \eprint{2401.02924}.

\bibitem[{\citenamefont{Ay\'on-Beato et~al.}(2015)\citenamefont{Ay\'on-Beato, Hassa\"\i{}ne, and M\'endez-Zavaleta}}]{Ayon-Beato:2015ada}
\bibinfo{author}{\bibfnamefont{E.}~\bibnamefont{Ay\'on-Beato}}, \bibinfo{author}{\bibfnamefont{M.}~\bibnamefont{Hassa\"\i{}ne}}, \bibnamefont{and} \bibinfo{author}{\bibfnamefont{J.~A.} \bibnamefont{M\'endez-Zavaleta}}, \bibinfo{journal}{Phys. Rev. D} \textbf{\bibinfo{volume}{92}}, \bibinfo{pages}{024048} (\bibinfo{year}{2015}), \bibinfo{note}{[Addendum: Phys.Rev.D 96, 049905 (2017)]}, \eprint{1506.02277}.

\bibitem[{\citenamefont{Tangen}(2007)}]{Tangen:2007yn}
\bibinfo{author}{\bibfnamefont{K.}~\bibnamefont{Tangen}} (\bibinfo{year}{2007}), \eprint{0705.4372}.

\bibitem[{\citenamefont{Lemos}(1995)}]{Lemos:1994xp}
\bibinfo{author}{\bibfnamefont{J.~P.~S.} \bibnamefont{Lemos}}, \bibinfo{journal}{Phys. Lett. B} \textbf{\bibinfo{volume}{353}}, \bibinfo{pages}{46} (\bibinfo{year}{1995}), \eprint{gr-qc/9404041}.

\end{thebibliography}

\end{document}